\documentstyle[epsfig]{mn}

\input{psfig.sty}

\title[Photo-Chemical Evolution of Elliptical Galaxies]
{Photo-Chemical Evolution of Elliptical Galaxies\\
I. The high-redshift formation scenario}

\author[A. Pipino and F. Matteucci]{Antonio Pipino,$^1$ 
and Francesca Matteucci,$^1$\\
$^1$Astronomy Department, University of Trieste,
    Via G.B. Tiepolo, 11, I-34127, Trieste, Italy}
\date{Accepted,
      Received }

\begin{document}
\maketitle

\begin{abstract}

In this paper we compute new multi-zone photo-chemical evolution models for 
elliptical galaxies, taking into account detailed nucleosynthetic yields, feedback from 
supernovae and an initial infall episode. By
comparing model predictions with observations, we derive a picture of galaxy formation
in which the higher is the mass of the galaxy, the shorter are the infall and the star formation
timescales. Therefore, in this scenario, the most massive objects are older than the
less massive ones, in the sense that larger galaxies stop forming stars at earlier times.
Each galaxy is created outside-in, i.e. the outermost regions
accrete gas, form stars and develop a galactic wind very quickly, compared to the central core in which the
star formation can last up to $\sim 1.3$ Gyr. In particular, we suggest that 
both the duration of the star formation
and the infall timescale decrease with galactic radius. 
In order to convert theoretical predictions into line-strength indices, 
different calibrations 
are adopted and discussed, focussing in particular on their dependence 
on the $\alpha-$enhancement.

By means of our model, we are able to
match the observed mass-metallicity and color-magnitude relations for 
the center
of the galaxies as well as to reproduce the overabundance of Mg 
relative to Fe, observed
in the nuclei of bright ellipticals, and its increase with galactic mass.
Furthermore, we find that the observed Ca underabundance relative 
to Mg can be real,
due to the non-neglibile contribution of type Ia SN to the
production of this element.
We predict metallicity and color gradients inside the galaxies which
are in good agreement with the mean value of the observed ones.

Finally, we conclude that models with Salpeter IMF are the best ones in 
reproducing the majority of the properties of ellipticals,
although a slightly flatter IMF seems to be required in order 
to explain the colors
of the most massive galaxies.

\end{abstract}


\section{Introduction}

Any model of galaxy evolution presented so far had to overcome the strong challenge
represented by the observational fact that elliptical galaxies show a remarkable 
uniformity in their photometric and chemical properties.
The first proposed scenario of elliptical  formation was the so-called
monolithic collapse scenario (Larson 1974; Matteucci $\&$ Tornambe', 1987; Arimoto
$\&$ Yoshii, 1987, AY; Chiosi $\&$ Carraro, 2002). In this framework,
ellipticals are assumed to have formed at high redshift as a result of a rapid collapse of a gas cloud.
This gas is then rapidly converted into stars by means of a very strong burst,
followed by a galactic wind powered by the energy injected into the interstellar medium (ISM) by
supernovae and stellar winds.
The wind carries out the residual gas from the galaxies, thus inhibiting further star formation.
Minor episodes of star formation, related to gas accretion from the surrounding medium
or interactions with neighbors, are not excluded although they do not influence the galactic evolution.
On the other hand, thanks to the success
of the cold dark matter theory in reproducing the anisotropy of the Cosmic Microwave Background, 
an alternative scenario of galaxy formation based on the Hierarchical Clustering scenario for the formation of dark matter halos was proposed.
Hierarchical semi-analytic models predict that
ellipticals are formed by several merging episodes
which trigger star-bursts
and regulate the chemical enrichment of the system (White $\&$ Rees, 1978). In this picture
massive ellipticals form at relatively low redshifts through major mergers between
spiral galaxies (e.g. Kauffmann $\&$ White, 1993; Kauffmann $\&$ Charlot, 1998). 

In the following we will refer to the monolithic collapse as the model in which ellipticals formed relatively quickly and at high redshift as opposed to the hierarchical clustering scenario where ellipticals are believed to form over a large redshift interval. 

The high redshift formation of
ellipticals is supported by observations showing an
increase in the strength of the metal absorption lines (Mass-Metallicity relation, e.g. Carollo et al.,
1993; Gonzalez, 1993; Davies et al., 1993; Trager et al. 1998, 2000)
and a reddening of the stellar light (Color-Magnitude relation, hereafter CMR,
Bower et al., 1992) with the velocity dispersion of the galaxies. If
we interpret the measure of the spectral indices as a measure of the
metallicity by means of a suitable calibration, we derive a picture in
which these two relations can be explained primarily as a
metallicity sequence (Kodama $\&$ Arimoto 1997). In other words, the most massive ellipticals are also the most metal rich. This fact was interpreted by Larson (1974) as due to the galactic
winds which should occur later in the most massive objects, thus allowing the star formation process to continue for a longer period than in small galaxies. In particular, the higher is the mass of the galaxy, the larger is
the chemical enrichment, the redder are
the colors. However, this interpretation based on the galactic winds has been changed in the last years, owing to the observed [Mg/Fe] ratio in the central parts of ellipticals and its trend with galactic mass.
In the last
decade there was an increasing evidence that [Mg/Fe] ratio is larger  than zero
in the core of bright galaxies (e.g. Faber et al., 1992; Carollo et
al., 1993; Davies et al., 1993; Worthey et al., 1992), suggesting
that the star formation lasted for a period shorter than the time at which the pollution from type 
Ia SNe becomes important (see Weiss et al., 1995). In fact, an overabundance of Mg relative 
to Fe is the clear sign that a short process of galaxy formation occurred before than a substantial number 
of type Ia SNe could explode and contribute to lower the [Mg/Fe] ratio (time-delay model, see Matteucci 
2001). In addition, the [Mg/Fe] ratio  in the cores of ellipticals increases with galactic mass 
(Worthey et al. 1992; Weiss et al. 1995; Kuntschner 2000, 2001) indicating, 
among the possible solutions, that the star formation lasted for a shorter period in the more massive 
systems ('inverse wind', Matteucci 1994). This is clearly at variance with the 'classical wind 
scenario' of Larson.
Other possible solutions to the increase on [Mg/Fe] as a function of galactic mass are a variable IMF 
and/or a decrease of the dark matter content as a function of galactic mass (see Matteucci et al., 1998).

On the other hand, recent models with selective merging
(i.e. large galaxies merge only with large galaxies) were able to reproduce
the Mg-$\sigma$ relation (e.g. Worthey $\&$ Collobert, 2003) and the CMR
(e.g. Kauffmann $\&$ Charlot, 1998), but they still
cannot explain the [Mg/Fe]-mass relation (Thomas et al. 1999, Thomas $\&$ Kauffmann 1999, Kuntschner 2000). 

Moreover, line-indices and color gradients are observed in ellipticals (e.g. Carollo et al., 1993; Peletier et al., 1990).
Also in this case, they can be reproduced by means of a variation in the
mean metallicity of the stellar population as a function of the radius (Tantalo et al., 1998b; Martinelli et al., 1998;
Tamura et al., 2000) in the context of monolithic models, whereas hierarchical assembly
of galaxies seems to produce milder slopes than the observed ones (e.g. Bekki $\&$ Shioya, 1999).

One of the main characteristics which makes ellipticals a very appealing test bench for
any scenario of galaxy formation, is that they exhibit tight relations
not only in the chemical properties, but also in the  dynamical ones.
It is well known that elliptical galaxies populate a plane (Fundamental Plane, FP) in the space
defined by the effective radius ($R_{eff}$), the mean surface brightness
within $R_{eff}$ and the central velocity dispersion $\sigma$ 
(Dressler et al., 1987; Djorgovski $\&$ Davies, 1987; 
Bender et al., 1992; Burstein et al. 1997). In particular, cluster ellipticals
show a very small scatter around the FP and such a tight scaling relation
is interpreted as an evidence of a highly standardized and synchronized formation of these galaxies.
Formation of ellipticals at high redshift can reasonably explain
the tightness of this plane, whereas predictions based on hierarchical models seem to be at odds with 
dynamical evidences deriving from the study of the FP, at least in the case of non-dissipative
merging (e.g. Nipoti et al., 2002, 2003). 
From a dynamical perspective, evidences favouring the hierarchical scenario are represented by the observed interacting 
galaxies (on-going mergers), ellipticals with disturbed morphologies
(i.e. counter-rotating cores, dust lanes, ripples, e.g. Kormendy $\&$ Djorgovski, 1989),
the morphology-density relation in clusters (Dressler et al., 1997) and the
Butcher-Oemler effect (Butcher $\&$ Oemler, 1978, although it could be related only to lenticular
galaxies).

Thanks to the new generation telescopes, the above-mentioned relations for the local universe 
are found to hold also at high redshift. In particular, 
recent observations of the lack of significant change in the slope and in the scatter of the CMR (Stanford et al., 1998),
the slow evolution of colors (e.g. Saglia et al. 2000; Ellis et al., 1997), mass to light
ratios (e.g. van Dokkum $\&$ Franx, 1996; Rusin et al., 2003; van de Ven et al., 2003) and line strength indices
of cluster early-type galaxies out to $z\sim 1$ (e.g. Bernardi et al., 2003; van Dokkum et al. 1998) 
suggest that their stars were formed at high redshift ($z>3$)
in a well synchronized epoch, and then evolved quiescently.
In summary, many observations (see Peebles 2002 for a review) suggest that ellipticals could already be in place
at $z\sim 2-3$, and their high-redshift counterpart could be represented by 
Lyman Break Galaxies (Steidel et al., 1996a,b)
which show chemical abundances and star formation rates consistent with those
of a rapidly evolving elliptical in their  early evolutionary phases (e.g. Matteucci $\&$ Pipino, 2002), or 
by Extremely Red Objects (EROs, e.g. Daddi et al., 2000; Miyazaki et al., 2003).

Peebles (2002) critically reviewed the still standing problems and the differences
between these two competing scenarios for galaxy formation, pointing out that this classification 
is now becoming meaningless, 
surviving only in historical sense (for another very recent review  
see de Freitas Pacheco et al., 2003),
since probably the best way to explain the whole set of observables within a cosmological
framework would request both scenarios to converge. 
Therefore, in order to shed light on this issue, 
this paper represents the first in a series in which we intend to investigate
the formation and the evolution of elliptical galaxies and the role 
played by the SN feedback, infall timescale, star formation efficiency
and stellar nucleosynthesis in driving the mass-metallicity and the
color-magnitude relations, as well as the Mg overabundance and the radial
gradients in metallicity and colors. 

In particular, in this article we present a new photo-chemical evolution 
model for
ellipticals, in which, for the first time, the inverse
wind scenario plus an initial infall episode are adopted whithin a multi-zone
formulation. The model results are then
compared with the observations, in order to fix the best set for our
input parameters and to define the best model.
Under reasonable assumptions on the behaviour of the infall timescale
and the star formation efficiency with galactic mass, we
show how this kind of model can reproduce
the whole set of chemical and photometric observables simultaneously.

In the forthcoming papers we shall compare
our best-model results with those from 
a hierarchical formation scenario, merger-induced starburst,
prolonged star formation, and we shall make
predictions for X-ray halos of ellipticals and the chemical enrichment of the 
intra-cluster medium. 
This paper is organized as follows: in section 2 we present the model. 
Section 3 is dedicated to the photo-chemical properties. In section 4 
we discuss
the results and draw the conclusions for our best model. 
In the following, conversion from cosmic time to redshift is done 
for a cosmological model with $\Omega_m=0.3$, $\Omega_{\Lambda}=0.7$
and $H_o=70 \rm \,km\,  s^{-1}\, Mpc^{-1}$. For all models we
assume a redshift of formation of 5. Under this assumption
we have an age of $\sim 12.3 \rm \, Gyr$ for our model galaxies. 
This age has been
chosen in order to reproduce the observed CMR.

\section{The model}

The adopted chemical evolution model
is an up-dated version of the multi-zone model of Martinelli et al. (1998) 
and Pipino et al. (2002),
in which we divide an elliptical galaxy into several non-interacting shells.
In particular, we use spherical shells with a fixed thickness of 0.1 $R_{eff}$, where $R_{eff}$ is the 
galactic effective radius, and limit our analysis to 10 $R_{eff}$, unless otherwise
stated. In the following we refer to the $0 - 0.1 R_{eff}$ as the central zone, or the
galactic core.
For each shell we calculate the evolution of the elemental abundances
by means of the equation of chemical evolution:
\begin{eqnarray*}
{d G_i (t) \over d t}  =  -\psi (t) X_i (t)\, + \\
                         +\int_{M_L}^{M_{B_m}} \psi (t-\tau_m) Q_{mi}(t-\tau_m) \phi (m) dm\, +\\
                         + A\int_{M_{B_m}}^{M_{B_M}} \phi (m) \left [ \int_{\mu_{min}}^{0.5} f(\mu) \psi (t-\tau_{m_2}) d\mu \right ] dm \, +\\
                         +(1-A)\, \int_{M_{B_m}}^{M_{B_M}} \psi (t-\tau_m) Q_{mi}(t-\tau_m) \phi (m) dm\, +\\
                         +\int_{M_{B_M}}^{M_U} \psi (t-\tau_m) Q_{mi}(t-\tau_m) \phi (m) dm\,  +\\
		         +({d G_i (t) \over d t})_{infall}\,  ,	
\end{eqnarray*}
where $G_i (t)= \rho_{gas}(t) \,X_i (t) / \rho_{gas}(0)$ is the mass density
of the element \emph{i} at the time \emph{t}
normalized to the initial value of the gas density. $X_i (t)$ 
is defined as the abundance by mass of the element \emph{i}.
By definition $\sum_i X_i=1$. We refer the reader
to Matteucci $\&$ Gibson (1995) and Matteucci $\&$ Greggio (1986) 
for a comprehensive discussion of this equation. Here we recall
that the integrals in the right-hand side of the equation give the 
rate at which the element \emph{i} is restored into the interstellar medium (ISM)
as unprocessed or newly-synthesized element by low- and intermediate-mass 
stars, SNIa and SNII, respectively. The last term represents the infall rate,
namely the rate at which the gas is assembling to form the galaxy. 

The variable $\psi$ is the star formation rate, for which we adopted the following law: 
\begin{equation}
\psi (t)= \nu\cdot \rho_{gas} (t)\, ,
\end{equation} 
namely it is assumed to be proportional to the gas density
via a constant $\nu$ which represents the star formation efficiency.
We assume $\nu$ as an increasing function of the galactic mass 
(see Tables 1, 2 and 3) in order to reproduce the 'inverse wind
model' (Matteucci, 1994, Matteucci et al., 1998).  The star formation
history of a particular shell is thus determined by the interplay
between the infall timescale at that radius, the star formation
efficiency and the occurrence of the galactic wind (i.e. the energetic
feedback from SNe and stellar winds).  In each zone we assume that
$\psi=0$ after the development of the wind.
 
$\phi (m)\propto m^{-(1+x)}$ is the initial mass function (IMF), normalized
to unity in the mass interval $0.1 -100 M_{\odot}$. In the following
we shall use only the exponents x=1.35 (Salpeter, 1955) and x=0.95 (AY).

One of the fundamental points upon which our model is based,
is the detailed calculation of the SN explosion rates.
For type Ia SNe we assume a progenitor model 
made of a C-O white dwarf plus a red giant (Whelan $\&$ Iben, 1973), then we have:
\begin{equation}
R_{SNIa}=A\int^{M_{BM}}_{M_{Bm}}{\varphi(M_B) \int^{0.5}_{\mu_m}{f(\mu)
\psi(t- \tau_{M_{2}})d \mu \, dM_{B}}}\, ,
\end{equation}
(Greggio $\&$ Renzini, 1983; Matteucci $\&$ Greggio, 1986), where $M_{\rm B}$ is the total mass of the
binary system, $M_{Bm}=3 M_{\odot}$ and $M_{BM}=16 M_{\odot}$ are the minimum
and maximum masses allowed for the adopted progenitor systems, respectively.
$\mu=M_2/M_{\rm B}$ is the mass fraction of the secondary, which
is assumed to follow the distribution law:
\begin{equation}
f(\mu)=2^{1+\gamma}(1+\gamma)\mu^\gamma\, .
\end{equation} 
Finally, $\mu_{m}$ is its minimum value and $\gamma=2$
The predicted type Ia SN explosion rate is constrained
to reproduce the present day observed value (Cappellaro et al., 1999), by
fixing the parameter $A$ in eq. (2). In particular, $A$ represents 
the fraction of binary systems in the IMF which are
able to give rise to SN Ia explosions. In the following we adopt $A=0.18\,(0.05)$ for a Salpeter (AY) IMF.

On the other hand, the type II SN rate is:
\begin{eqnarray}
R_{SNII} & = & (1-A)\int^{16}_{8}{\psi(t-\tau_m) \varphi(m)dm}\nonumber \\
& + & \int^{M_U}_{16}{\psi(t-\tau_m) \varphi(m)dm}\, ,
\end{eqnarray}
where the first integral accounts for the single stars in the 
range 8-16$M_{\odot}$, and $M_{U}$ 
is the upper mass limit in the IMF.

\subsection{Energetics}

\subsubsection{Binding Energy}
We evaluated the binding energy of the gas in the \emph{i-th} zone as

\begin{equation}
E_{Bin}^i (t) = \int_{R_i}^{R_{i+1}} d L(R)\, ,
\end{equation}
where $d L(R)$ is the work required 
to carry a quantity $dm= 4\pi R^2 \rho_{gas} dR$
of mass out to infinity and ${R_i}$ is the radius of the \emph{i-th} shell
(Martinelli et al., 1998). The baryonic matter (i.e. star plus gas)
is assumed to follow the distribution (Jaffe, 1983):

\begin{equation}
F_l (r)\propto {r/r_o \over 1+r/r_o }\, , 
\end{equation}
where $r_o = R_{eff}/0.763$.

We assume that the dark matter (DM) is distributed in a diffuse halo
ten times more massive than the baryonic component of the galaxy
with a scale radius $R_{dark}= 10 R_{eff}$ (Matteucci 1992), where $R_{eff}$
is the effective radius. The DM profile is taken from
Bertin et al. (1992).
Other cases with a more concentrated
DM lead to delayed winds and thus to prolonged star formation, at odds
with observations (Martinelli et al. 1998).

\subsubsection{Feedback from Supernovae}
We adopt two different recipes for SNIa and SNII, respectively.
For SNII we assume that the evolution of the energy in the 'snowplow'
phase is regulated by Cioffi et al. (1988) cooling time. 
This timescale takes into account the effect of the metallicity of the gas, and
therefore, one can model in a self-consistent way the evolution of a SNR in
the ISM.  Due to significant cooling by metal ions, only a few percent
of the initial $\sim 10^{51}\rm\, erg$ (blast wave energy) can be provided to the ISM.  On
the other hand, for type Ia SNe, we follow the suggestions put forward
by Recchi et al. (2001) in the modelling of dwarf galaxies. Recchi et al. (2001)
assume that, since SNIa explosions occur in a medium already heated by
SNII, they contribute with the total amount of their energy budget
($10^{51}\rm\, erg$), without radiative losses. This result has been already
adopted by Pipino et al. (2002), who reproduced realistic
models for ellipticals and the Fe abundance in the ICM.  However, Pipino et al. (2002)
stressed that, owing to the larger number of SNII relative to SNIa, the
typical efficiency of energy release to the ISM averaged on both types of SN is $\sim$ 20\%.
Since we use a chemical evolution code which 
adopts the same formulation for the feedback as in Pipino et al. (2002),
we consider a $\sim$ 20\% mean efficiency in energy transfer as a  
representative value also for the model galaxies presented in this paper.

We define the time when the galactic wind occurs ($t_{gw}$) as the solution of
the following equation:

\begin{equation}
E_{th}^i(t_{gw}) = E_{Bin}^i(t_{gw})\, ,
\end{equation}


\subsection{Infall}

The main novelty of this paper relative to our previous ones (Matteucci et al., 1998; Martinelli et al., 1998;
Pipino et al., 2002) is that we simulate the creation of the spheroid as
due to the collapse of either a big gas cloud or several smaller gas lumps. 
The infall term in the chemical modelling of ellipticals was first introduced by
Tantalo et al. (1996) and Kodama $\&$ Arimoto (1997) 
following a suggestion of Bressan et al. (1994),
in order to solve a problem similar to the 'Classical G-Dwarfs Problem'
(e.g. Tinsley 1980) in the Milky Way. In particular, Bressan et al. (1994; see
also Worthey et al., 1996)
claimed that the UV-excess predicted by their theoretical spectral energy 
distribution (SED) could be reconciled with the observed one only
by reducing the excess of low-metal stars. This can
be done by relaxing the closed-box approximation and introducing an initial 
infall of gas which modulates the star formation, thus preventing
the formation of too many low metallicity stars. The infall makes the star formation rate
to start from a smaller value than in the closed box case,
to reach a maximum and, then, to decrease when the star formation process becomes dominant.
This treatment is certainly more realistic since it includes the simulation of a gas collapse.

The infall term is present in the right-hand side of the equation of the chemical 
evolution, and the adopted expression is:
\begin{equation}
({d G_i (t) \over d t})_{infall}= X_{i,infall} C e^{-{t \over \tau}}\, ,
\end{equation}
where $X_{i,infall}$ describes the chemical composition of the accreted gas, assumed to be primordial.
$C$ is a constant obtained by integrating the infall law over time and
requiring that $\sim 90\%$ of the initial gas has been accreted at $t_{gw}$ (in fact,
we halt the infall of the gas at the occurrence of the galactic wind).
Finally, $\tau$ is the infall timescale. 
We note that models in which
$\tau \rightarrow 0$ are equivalent to the standard closed box model
with all the gas already in place when the star formation starts.

A novelty in our formulation is that we adopted a new
approach to the gas accretion history. In fact, at variance with previous works 
where $\tau$ was a-priori related to the free-fall timescale (Tantalo et al., 1996;
Romano et al., 2002), we decided to consider 
the infall timescale as a free parameter. 
In particular, we decided to tune $\tau$ in each zone of the galaxy by requiring our models to
fit the metallicity and color gradients (for a detailed
discussion see section 3.2), as well as the mass metallicity and the color-magnitude ralations.
In the following we show results for models with
infall timescales assumed to be either constant or a decreasing function of the mass.
On the other hand, we adopted in each model a star formation efficiency increasing with mass, in order
to predict a [Mg/Fe] ratio increasing with galactic mass, as suggested by Matteucci
(1994).

\subsection{Stellar Yields}

We follow in detail the evolution of 21 chemical elements
and to do that we need to adopt specific prescriptions for the stellar nucleosynthesis.
In particular, our nucleosynthesis prescriptions are:
\begin{enumerate}
\item
For single low and intermediate mass stars ($0.8 \le M/M_{\odot} \le 8$) we make use of the yields 
of  van den Hoek $\&$ Groenewegen (1997) as a function of metallicity. 
\item
We use the yields by Nomoto et al. (1997) for SNIa which are assumed to originate from C-O white dwarfs
in binary systems which accrete  material from a companion (the secondary) and reach the Chandrasekar 
mass and explode via C-deflagration . 

\item
Finally, for massive stars ($M >8 M_{\odot}$) we adopt the yields of  Thielemann et al. (1996, TNH96)
which refer to the solar chemical composition, 
and the yields of
Woosley $\&$ Weaver (1995, WW95 hereafter) corresponding to the solar chemical composition. 
In particular, we assumed WW95 case A for stars with masses $<30M_{\odot}$, and case B for $\ge 30M_{\odot}$. 
We also computed one case where the yields of Mg of WW95 were corrected as recently suggested by 
Fran\c cois et al. (2003) (case WW95m) to best fit the solar neighbourhood stars.
In particular, we increased the contribution to the Mg by a factor of $\sim 10$ in the mass range
$11-22M_{\odot}$ and lowered it by the same factor in stars with mass $>22 M_{\odot}$, 
relative to the original
WW95 results. 
In the following we always refer to the TNH96 yields, unless otherwise stated.
 
\end{enumerate}

\subsection{Photometry}

We calculate the detailed photometric evolution for elliptical galaxies of different baryonic mass
($10^{10}$, $10^{11}$, $10^{12}M_{\odot}$) by applying
the spectro-photometric code by Jimenez et al. (1998), where all the details can be found. 
In particular, we reconstruct the composite stellar population
(CSP) inhabiting  each zone of the galaxies, for which we know 
the star formation and chemical evolution history.
In tables 1, 2 and 3 we show the predicted K- and B-band luminosities of the galaxy models,
obtained by integrating the contributions of each zone out to 10 $R_{eff}$ 
whereas the colors for the inner zone are plotted in Fig. 10.

\subsection{Galactic models}

We run models for elliptical galaxies in the baryonic mass range 
$10^{10}-10^{12}M_{\odot}$.
$M_{lum}$ is the 'nominal' mass of the object, i.e. the mass of the initial gas budget (we recall that we normalize
the infall law between $t=0$ and $t\sim t_{gw}$). The mass in stars at the present time
is $\sim 0.2-0.4$ $M_{lum}$ for all the models and the velocity dispersion $\sigma$ is evaluated
from the relation $M=4.65\cdot 10^5\, \sigma^2\, R_{eff}\, M_{\odot}$ (Burstein et al., 1997). 
The effective radius $R_{eff}$ is the final one, achieved when the collapse is over and this occurs
roughly at the time $\tau$.

The models are:

\begin{enumerate}
\item Model I:  Salpeter IMF, $\tau$ constant with galactic mass (see Table 1).
\item Model II: Salpeter IMF, $\tau$ decreasing with galactic mass (see Table 2).
\item Model III: as Model II, but with Arimoto $\&$ Yoshii IMF (see Table 3).
\end{enumerate}

Moreover, for each model we considered the following cases:

\begin{enumerate}
\item[a:] $\tau$ constant with radius. 
\item[b:] $\tau$ decreasing with radius.
\item[c:] $\tau$ increasing with radius.
\end{enumerate}

\begin{table*}
\centering
\begin{minipage}{120mm}
\scriptsize
\begin{flushleft}
\caption[]{Model I}
\begin{tabular}{l|llllllll}
\hline
\hline
$M_{lum}$ 	&$R_{eff}$ &  $\nu$ 	    & $\tau$& $t_{gw}$ & $Mg_2$ & $[<Mg/Fe>]$&$L_B$ &$L_K$ \\
({$M_{\odot}$}) & ({kpc})  &  ({$Gyr^{-1}$})& (Gyr)& (Gyr)   &        &              &($10^{10}L_{\odot}$)&($10^{10}L_{\odot}$)\\
\hline
$10^{10}$       & 1	   & 3	            & 0.2    & 1.114   & 0.276  & 0.259     & 0.08 &0.35 \\
$10^{11}$       & 3        & 10             & 0.2   &  0.570  & 0.283  & 0.305     & 0.79 &3.95\\
$10^{12}$       & 10       & 27             & 0.2   &  0.499  & 0.327  & 0.316     & 7.25 &40.6	\\
\hline
\end{tabular}
The values presented in this table are referred to the central ($0 \rightarrow 0.1 R_{eff}$) zone of the models.
The calibration used for $Mg_2$ is from Matteucci et al. (1998), see section 3. The luminosities are referred to the whole galaxy.
\end{flushleft}
\end{minipage}
\end{table*}

\begin{table*}
\centering
\begin{minipage}{120mm}
\scriptsize
\begin{flushleft}
\caption[]{Model II}
\begin{tabular}{l|llllllll}
\hline
\hline
$M_{lum}$ 	&$R_{eff}$ &  $\nu$ 	    & $\tau$& $t_{gw}$ & $Mg_2$ & $[<Mg/Fe>]$&$L_B$ &$L_K$\\
({$M_{\odot}$}) & ({kpc})  &  ({$Gyr^{-1}$})& (Gyr)& (Gyr)   &        &              &($10^{10}L_{\odot}$)&($10^{10}L_{\odot}$)\\
\hline
$10^{10}$       & 1        & 3              & 0.5  & 1.299   &  0.267  & 0.150	  &0.07	&0.33 \\
$10^{11}$       & 3        & 10             & 0.4   &  0.709  & 0.298  & 0.206     & 0.78 &3.92\\
$10^{12}$       & 10       & 22             &0.2    &   0.544  & 0.328  & 0.303     & 7.42 &41.3\\
\hline
\end{tabular}
\end{flushleft}
The values presented in this table are referred to the central ($0\rightarrow 0.1 R_{eff}$) zone of the models.
The calibration used for $Mg_2$ is from Matteucci et al. (1998), see section 3. The luminosities are referred to the whole galaxy.
\end{minipage}
\end{table*}

\begin{table*}
\centering
\begin{minipage}{120mm}
\scriptsize
\begin{flushleft}
\caption[]{Model III}
\begin{tabular}{l|llllllll}
\hline
\hline
$M_{lum}$ 	&$R_{eff}$ &  $\nu$ 	    & $\tau$& $t_{gw}$ & $Mg_2$ & $[<Mg/Fe>]$&$L_B$ &$L_K$\\
({$M_{\odot}$}) & ({kpc})  &  ({$Gyr^{-1}$})& (Gyr)& (Gyr)   &        &              &($10^{10}L_{\odot}$)&($10^{10}L_{\odot}$)\\
\hline
$10^{10}$       & 1        & 5              & 0.3   &1.234   & 0.421  & 0.534     & 0.03&0.39\\
$10^{11}$       & 3        & 14             & 0.2   & 0.650   & 0.429  & 0.535     & 0.38 & 4.66\\
$10^{12}$       & 10       & 27             & 0.1   &0.504   & 0.453  & 0.581     & 4.00 &50.0\\
\hline
\end{tabular}
\end{flushleft}
The values presented in this table refer to the central ($0\rightarrow 0.1 R_{eff}$) zone of the models.
The calibration used for $Mg_2$ is from Matteucci et al. (1998), see section 3. 
The luminosities refer to the whole galaxy.
\end{minipage}
\end{table*}

\section{Results}

\subsection{The [el/Fe] vs. [Fe/H] relations}

First of all we present the chemical properties predicted in the core of ellipticals.
In particular, we show the behaviours of the main chemical species in the ISM of 
these galaxies. 
Unfortunately, the only way to derive abundances in ellipticals is through metallicity indices
measured in integrated stellar spectra. 
Therefore, in order to compare 
our results with observations, we need to transform our model predictions into
line-strength indices (see section 3.1.1), thus adding possible sources of systematic errors. 
In the following we present our results concerning the real abundances expressed in the usual \emph{bracket} notation, where
solar abundances are taken from Anders $\&$ Grevesse (1989), unless otherwise stated.

The main parameters of Models I, II and III , as well as their predicted photo-chemical 
properties are shown in Tables 1, 2 and 3, respectively. In particular, in column 1 we present the 
luminous mass of the model.
The assumed effective radius,
star formation efficiency, infall timescale and galactic wind occurrence for the 
central (i.e. $0-0.1 R_{eff}$) zone
are shown in columns 2, 3, 4 and 5 respectively. In column 6 we show
the $Mg_2$ index predicted by means of Tantalo et al.'s SSP (see next section), whereas
in column 7 we present the predicted mean stellar $\rm[<Mg/Fe>]_{*}$. The last two columns 
contain the predicted B- and K-band galactic luminosities.
We stress that the values in the galactic core are constrained by the observed relations,
so are the same in all the different cases (a, b and c). In other words, the results
presented in these Tables are indipendent of the assumed relation between $\tau$ and the galactic radius.
As expected from the assumed increase of $\nu$ with the galactic mass, 
in each model the galactic wind occurs earlier in the more massive galaxies. It is worth noting that in 
Larson's model (1974), as well as in Arimoto $\&$ Yoshii's (1987) and Matteucci $\&$ Tornamb\`e's (1987) models, the
efficiency of star formation was assumed to be constant or to decrease with galactic mass, 
thus leading to 
the situation of the classical wind scenario. 
In the core of the largest objects the active evolution stops  at $\sim 0.5$ Gyr,
whereas it can last more than 1 Gyr in the central regions of a $10^{10}M_{\odot}$ galaxy.
As we shall see in section 3.1.2, this sequence in the development of the galactic winds
is needed to reproduce the observed Mg enhancement relative to Fe in bright ellipticals.
In our models $\tau$ and $\nu$ work together in the same direction, 
producing, in the more massive systems,
a star formation rate 
peaked at earlier times and with higher peak values than in the smaller ones. 
The increase of the star formation efficiency, in spite of the shorter duration of the star 
formation process, 
leads to a higher metal enrichment
in the most massive galaxies. Therefore, this mechanism preserves the mass-metallicity relation
which was the main achievement of the classic wind models.

In Fig. 1 we plot the predicted curves for the [$\alpha$/Fe] abundance ratios versus [Fe/H]
in the ISM of the central zone ($0-0.1 R_{eff}$) for a
$10^{11}M_{\odot}$ Model II galaxy. Similar trends are predicted by Model I,  
whereas we obtain systematically higher [$\alpha$/Fe] ratios 
in the curves predicted by Model III, due
to its flatter IMF. The high values ([$\alpha$/Fe]$>1$) at very low metallicity
are due to the adopted yields, as it can be clearly seen from Fig. 2 comparing Model II
predictions (dotted curve) with those from Pipino et al. (2002) best model, adopting the yields
of Woosley $\&$ Weaver (1995), which are
systematically lower by $\sim 0.2-0.4$ dex. On the other hand, the $\alpha$-element 
overabundance 
relative to Fe,
and its decrease with increasing metallicity is
due to the different origin of these elements (time-delay model, Matteucci $\&$ Greggio, 1986, 
see section 2.3). 

The $\alpha$-elements exhibit different degrees of enhancement
with respect to Fe. This is due, from a theoretical point of view, to the different degree of production of each element in type II and Ia SNe. In particular, Si and Ca show a lower overabundance relative to Fe than O and Mg.
From an observational point of view, the measure
of the CaII triplet in the near infrared is considered as a good tracer for the global metallicity, since
it is independent from the age (e.g. Idiart et al. 1997). The observations show that, 
although Ca belongs to the $\alpha$ elements, the strength of the observed lines
follows very closely [Fe/H] instead of [Mg/H] (Worthey, 1998; Trager et al., 1998; Saglia et al. 2003). 
In particular Thomas et al. (2003) suggested [Ca/Mg] = -0.3, and Saglia et al. (2003), after a detailed 
analysis of several possible sources
of error, claimed that the Ca depletion in ellipticals might be real. 
Our calculations suggest
[$\rm<Ca/Mg>_*$] =-0.152 and [$\rm<Ca/Fe>_*$] =-0.03 (these values are referred to the core of a $10^{11}M_{\odot}$ Model II galaxy).
These results can be explained simply in terms of yields, since Ca is produced in a non-negligible
amount by type Ia SNe.
In fact, the mass of Ca ejected during a SNIa explosion in the model W7 (Nomoto et al., 1997) 
is $\sim 0.012 M_{\odot}$, whereas the contribution of type II SNe averaged on a Salpeter IMF
in the mass range $10 - 50 M_{\odot}$ is $\sim 0.0058 M_{\odot}$ (see 
Table 3 of Iwamoto et al., 1999).
Silicon, for the same reason of Ca,  
presents a very similar behaviour and a clear evidence of these departures from the 
$\alpha$ element abundance pattern (represented by O and Mg) at high [Fe/H], as it can be
be seen in Fig. 1.

\begin{figure}
\epsfig{file=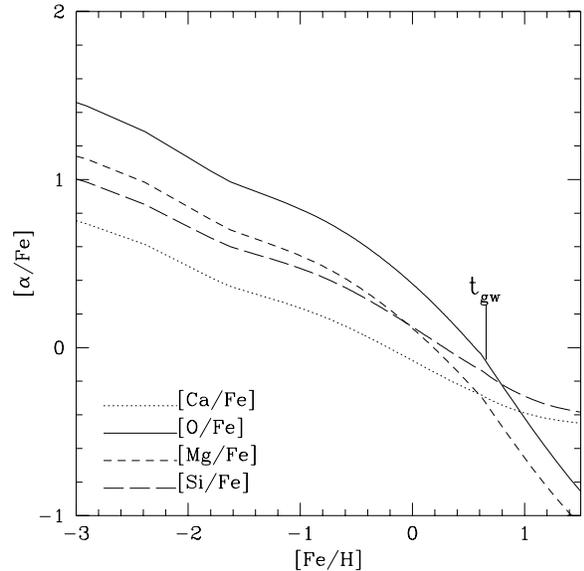, height=8cm, width=8cm}
\caption{Theoretical [O, Mg, Si, Ca/Fe] abundance ratios in the ISM as functions of [Fe/H]
predicted by Model II, for the core
of a $10^{11}M_{\odot}$ galaxy. Solar value for O by Holweger (2001). The time for the occurrence of the galactic wind, $t_{gw}$,  is indicated.}
\end{figure}

\begin{figure}
\epsfig{file=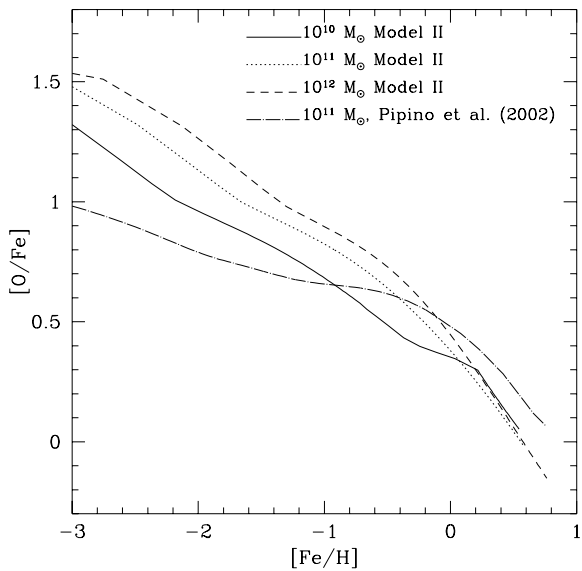,height=8cm,width=8cm}
\caption{Theoretical [O/Fe] abundance ratios in the ISM as functions of [Fe/H]
predicted by Model II for the core
of $10^{10}-10^{12}M_{\odot}$ galaxies. Solar value for O by Holweger (2001).
Results for Pipino et al. (2002) best model galaxy of $10^{11}M_{\odot}$ are shown by a
long-dashed-dotted line. 
}
\end{figure}

The behaviour of the [O/Fe] abundance ratio in the ISM of several galactic models is analysed in Fig.
2, where we plot the new model results for different galactic masses
compared to predictions from Pipino et al. (2002) best model. 
From Fig. 2 we infer a higher [O/Fe] in the ISM of the larger galaxies at a fixed [Fe/H], as
a consequence of the effect of the more efficient star formation rate in the brightest galaxies
relative to the smaller ones.

\begin{figure}
\epsfig{file=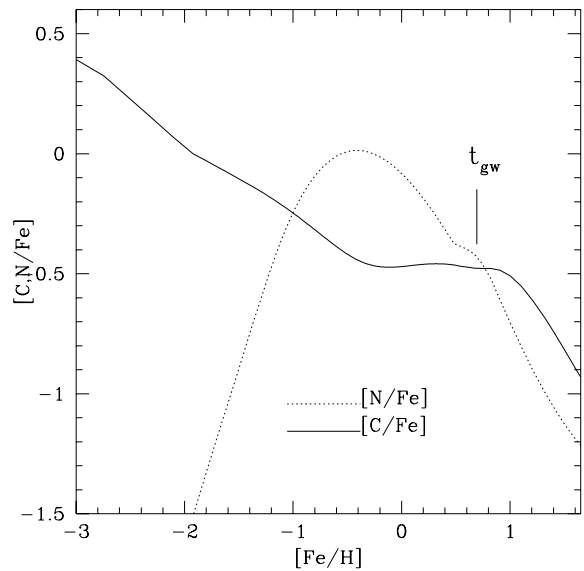,height=8cm,width=8cm}
\caption{Theoretical [C,N/Fe] abundance ratios in the ISM as functions of [Fe/H]
predicted by Model II for the core of a $10^{11}M_{\odot}$ galaxy.}
\end{figure}

Finally, in Fig. 3, theoretical [C/Fe] and [N/Fe] abundance ratios in the ISM as functions of [Fe/H]
are shown for the core of a $10^{11}M_{\odot}$ Model II galaxy. The fast increase
of N abundance with metallicity is expected, since N produced in massive stars is assumed to be only secondary (but see recent calculations of Meynet $\&$ Maeder 2002, where a small fraction of primary N from massive stars is predicted). On the other hand, some primary N is produced in intermediate mass stars in the yields adopted here.
C exhibits a monothonic decrease
that parallels the [O/Fe] behaviour although the predicted overabundance of C relative to Fe is much lower than for the $\alpha$-elements and this is due to the fact that C is largely produced in low and intermediate mass stars.

\subsubsection{Mass-Metallicity relations for ellipticals}

From the observational point of view, we infer the presence of the so-called
mass-metallicity relation from the increasing strength of the metallic absorption lines
with increasing velocity dispersion of the galaxy (e.g. Faber et al., 1977; Carollo et al.,
1993; Gonzalez, 1993; Davies et al., 1993; Trager et al. 1998, 2000). 
At the present time, these relations are well established thanks to the increasing 
number of observed galaxies (e.g. Colless et al., 1999, give $\rm Mg_{2}= 0.257 log \sigma -0.305$, 
with an intrinsic scatter of 0.016 mag. On the other hand, Bernardi et al., 2003,   
with their huge sample of early type galaxies at $z<0.3$,  
found $\rm log Mg_{2}\simeq (0.20\pm 0.02) log \sigma $, with an intrinsic scatter of 0.011 mag,
and $\rm log <Fe>\simeq (0.11\pm 0.03) log \sigma $, with a scatter of 0.011). 
 
In the following we focus
on the two indices $Mg_2$ and $<Fe>$, which are related to the metallicity of the observed object,
but carry also information on its age (because of the well-known age-metallicity degeneracy, e.g. 
O'Connel, 1976). 

The first step in order to compare our predicted abundances with the 
observed indices, is to compute the mean stellar abundance of the element X ($<X/H>\equiv <Z_X>$),
defined as (Pagel $\&$ Patchett 1975):
\begin{equation}
<Z_X>={1\over S_0} \int_0^{S_0} Z_X (S) dS\, ,
\end{equation}
where $S_0$ is the total mass of stars ever born contributing to the 
light at the present time. We recall that, for massive ellipticals, 
results obtained by averaging on the stellar mass are very similar 
to those obtained by
averaging on the stellar luminosity at the present time (which
is the right procedure, since the observed indices are weighted
on V-band luminosity, see e.g. Arimoto $\&$ Yoshii, 1987; Matteucci et al., 1998).
Tests on the $10^{11}M_{\odot}$ Model II galaxy confirm these findings.
In particular, the differences between these two alternatives
result into an offset of only a few percents in the predicted indices. 
Then, we transform our chemical information into indices
by means of a \emph{calibration} relation.
We adopt the calibrations derived by Matteucci et al. (1998) from the synthetic indices 
of Tantalo et al. (1998a), which takes into account the Mg
enhancement relative to Fe:

\begin{eqnarray*}
Mg_2=0.233+0.217[Mg/Fe]\\
+(0.153+0.120 [Mg/Fe])\cdot [Fe/H] \,  ,\\ 
<Fe>=3.078+0.341[Mg/Fe]\\
+(1.654-0.307 [Mg/Fe])\cdot [Fe/H] \,  , 
\end{eqnarray*}
for a stellar population 15 Gyr old. For comparative purposes
we use also the calibration relations of Worthey (1994) for a 12 Gyr old SSP with solar abundance ratios
and $[Fe/H]>-0.5$.

These relations are:
\begin{eqnarray*}
Mg_2= 0.187\cdot [Fe/H]+0.263\, ,\\
<Fe>=1.74 \cdot [Fe/H]+2.97\, ,
\end{eqnarray*}
In this way we can have a good estimate of the impact of $\alpha -$enhancement
in the predicted metallicity indices, and we are able to draw conclusions which are 
independent from the assumed calibration.

\begin{figure}
\epsfig{file=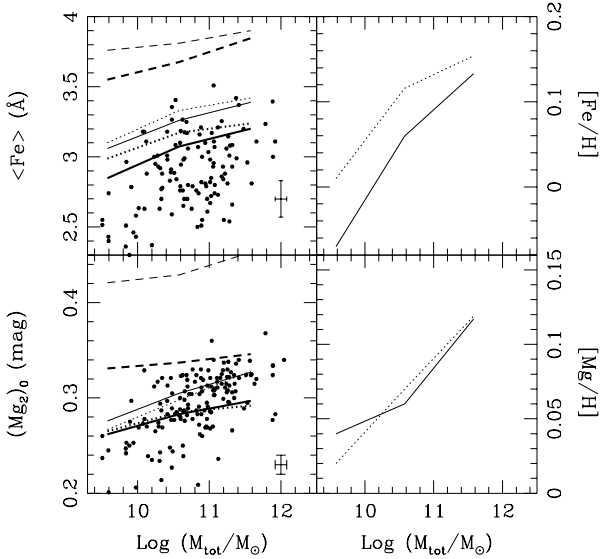,height=8cm,width=8cm}
\caption{\emph{Right}: Mg and Fe abundances in the stellar component predicted by Model I (solid) and Model II (dotted)
as functions of galactic mass.
\emph{Left}: Line-strength indices predicted by Model I (solid), Model II (dotted) and Model III (dashed),
using Tantalo et al. (1998a) and Worthey (1994) calibrations (thick and thin lines, respectively), plotted
versus a collection of data from Carollo et al. (1993), Trager et al. (1998), Gonzalez (1993), Kuntschner (2000), Kuntschner et al. (2001).
The typical errors are shown in the panels.\label{fig1}}
\end{figure}

Our models with Salpeter IMF (solid and dotted curves in Fig. 4, representing Model I and II, respectively)
fit reasonably well the observed mass-metallicity relations,  although
they predict a sligthly flatter slope with respect to the observations. 
On the contrary, the model with AY IMF (dashed line)
predicts too high values for the line-strength indices. 
The difference between the predicted slopes and the observed ones
could be due to a too low predicted metallicity in our high mass models, which reflects
in a lower $Mg_{2}$ than the observed one, but it could also depend on other factors
including the difference between the \emph{assumed} mass of the objects and that derived from the \emph{observed}
$\sigma$. Finally, it could be due to the adopted calibrations. 
In particular, from the bottom-left panel of
Fig. 4, we note that the difference between Model I and II predictions for $Mg_{2}$, for a fixed
calibration relation,
reflects in an offset of $<0.02$ mag, being the slope of the mass-metallicity relations nearly the same.

On the other hand, the slopes are different when considering the two calibrations, and the predictions
based on $\alpha -$enhanced tracks are systematically higher by $\sim 0.02-0.06$mag.
Going to numerical details, Model I predicts that Mg and Fe abundances variation (between the highest and the
lowest mass model) are $\rm \Delta [Mg/H]= 0.097\rm dex$ (being oversolar in both cases) and $\Delta [Fe/H]= 0.203 dex$ (see right-hand
panels of Fig. 4). 
These results translate into a $\rm \Delta Mg_2$ = 0.046 (Tantalo et al. calibration), $\rm \Delta Mg_2$ = 0.035 (Worthey calibration)
and $\rm \Delta log <Fe> \simeq 0.05$ (for both calibrations). For Model II we obtain very similar results, the only differences
being a somewhat flatter behaviour of [Fe/H] ($\rm \Delta [Fe/H]= 0.153 dex$) and a larger 
$\rm \Delta Mg_2$ (= 0.061, Tantalo et al. calibration), whereas $\rm \Delta Mg_2$ = 0.027 for the Worthey calibration.
The main driver of this difference arises from the variation of the [Mg/Fe] ratio, which increases faster in Model II than in Model I.
This implies that our model predictions are not independent from the assumed calibration, but this fact
is in general less important than a change in the IMF, as we can deduce by comparing 
the curves in Fig. 4 for Model III with the observed points. 
In fact, we rule out model III (AY IMF), whose predictions do not
match the observed indices (Fig. 4) and since the entries in 
column 7 of Table 3 (to be compared with the expected [Mg/Fe]$\sim 0.2-0.4$ of Fig. 6) show a too high Mg 
overabundance.
Model III fails in reproducing these quantities even if we assume an extremely short
infall timescale and WW95 yields for massive stars, which
give lower [Mg/Fe]. These results reinforce
the need of a Salpeter IMF in order to reproduce the whole set of observables.

For what concerns the mass-$<Fe>$ relation (always in Fig. 4), the lowest mass models predict in general a $\sim 0.2 - 0.3
\rm \AA$ higher $<Fe>$ index 
than the average observed value, also when the Worthey calibration is considered.
This disagreement could be either due to the adopted stellar yields,
or to uncertainties in the data
(see the large spread in the observations in Fig. 4, and in particular in the  $<Fe>$ index measurements which suffer
larger uncertainties than those of $Mg_{2}$. Typical values are $>0.2 \rm \AA$).

Finally, in the right panels of Fig. 4 we show our predicted Fe-mass and Mg-mass relations, when real abundances and not indices are considered. The Figure indicates a slight flattening of the [Fe/H] for the most massive ellipticals, which is not present in the predicted [Mg/H].
This can be due to the fact that in the inverse wind model picture larger galaxies develop galactic winds before the less massive ones and therefore the type Ia SNe have less time to restore Fe into the ISM.

\subsubsection{[Mg/Fe]} 

A particular emphasis should be given to 
the observed increase in the enhancement of Mg with respect to Fe going to massive ellipticals 
(e.g. Worthey et al., 1992;
Matteucci, 1994; Thomas, et al. 2002), since it represents a very strong constraint on 
their star formation history. 
One can envisage several possibilities to account for this trend: a shorter timescale of star formation which can be obtained by increasing the star formation efficiency with galactic mass which, in turn, induces the occurrence of galactic winds, with consequent loss of the residual gas,
earlier in the most massive objects 
(inverse wind picture, see Matteucci 1994).
Another possibility is to assume that the IMF varies from galaxy to galaxy becoming flatter with increasing galactic mass. Finally, a decreasing amount of dark matter with  increasing luminous mass or a selective loss of metals could be the cause for the increase of [Mg/Fe] (see Matteucci et al. 1998 for an extensive discussion).
In this paper we consider only the first possibility, as already discussed in the previous sections.

\begin{figure}
\epsfig{file=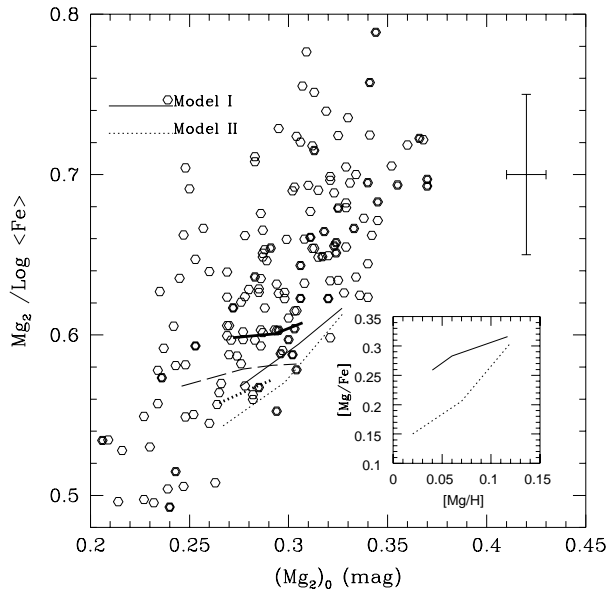,height=8cm,width=8cm}
\caption{Predicted relations for Model I (solid), Model II (dotted) and a model with \emph{classic}
wind (Pipino et al., 2002, dashed) are shown. Indices obtained with the Tantalo et al. calibration are
displayed with thick lines, whereas results 
obtained by means of the Worthey calibration are shown with thin curves.
Data from Gonzalez (1993), Carollo et al. (1993), Kuntschner (2000), 
Kuntschner et al. (2001). The typical error is shown in the top-right corner.
Mean stellar [Mg/Fe] versus [Mg/H] from the inner zones of Model I and II are shown in the small
figure. \label{fig2}}
\end{figure}

\begin{figure}
\epsfig{file=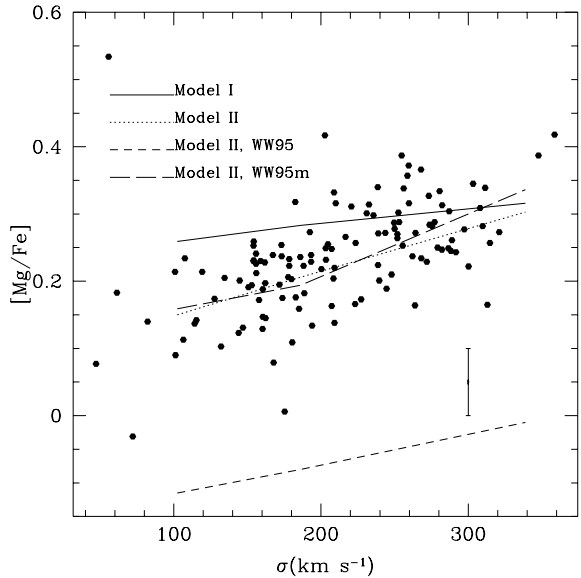,height=8cm,width=8cm}
\caption{[Mg/Fe] as a function of galactic velocity dispersion predicted by Model I (solid)
and II (dotted) compared to the
data from Gonzalez (1993), Mehlert et al. (2000), Beuing et al. (2002)
converted to $\alpha$/Fe ratios by Thomas, Maraston, $\&$ Bender (2002). 
The typical error is shown in the bottom-right corner. For comparison we show the
theoretical curves obtained with the same input parameters of Model II, but different yields (see text).
WW95: yields by Woosley $\&$ Weaver (1995). WW95m: yields by Woosley $\&$ Weaver (1995) modified according to 
Fran\c cois et al. (2003).\label{fig3}}
\end{figure}

In Fig. 5 we show the observed trend of Mg/Fe in elliptical galaxies by means
of the quantity $Mg_2 /log <Fe>$ plotted versus $(Mg_{2})_0$ (as a mass tracer) and our 
model curves.
We recall here that $Mg_2 $ is defined in magnitudes, whereas $<Fe>$ is in $\rm \AA$, so
in order to have consistent variations (i.e. on the same scale) of the two indices, we
decided to use $log <Fe>$ as a tracer for [Fe/H]. Although there is a good agreement between the predicted
and the observed slopes, our curves lie under the mean relation expected from the fit to the observed values. This 
is a consequence of the sligthly higher $<Fe>$ predicted by our
models (see top-left panel of Fig. 4 and related discussion).
We emphasize that the inverse wind model 
is necessary to reproduce the observed slope in the relations involving $\alpha$-elements 
and Fe. In fact, we see that the trend for $Mg_2 /log <Fe> vs. (Mg_{2})_0$ relation predicted by a 
\emph{classic} wind model (dashed line in Fig. 5), with similar input parameters 
(see also Pipino et al. 2002 for the details of this model),
is at variance with the observed behaviour. Note, however, the large uncertainty affecting
this kind of diagnostic. In the small panel we show for comparison the predicted trends
for the \emph{real} abundances in the stellar populations of our galactic models.

In Fig. 6 we present our predictions for the mean stellar [Mg/Fe] plotted against a
set of data analysed and transformed into [$\alpha$/Fe] ratios by Thomas et al. (2002).
It is clear that Model I does not predict the right slope for the [Mg/Fe]-$\sigma$
relation, and the results become worse if we use a smaller infall timescale. On
the other side, Model II reproduces fairly well the observed relation,   
thus leading us to the conclusion that it can be chosen as the best one in
our parameters space. With this choice we fixed the free parameters $\nu$ and $\tau$
in the nuclei of ellipticals. The other degrees of freedom, represented e.g. by
a possible change of $\nu$ and $\tau$ with radius (see section 2.5 for the definition
of the sub-models studied), will be constrained by
the analysis of the radial gradients in metallicity and colors 
(sections 3.2 and 3.4, respectively).

Finally, we note that our results may be affected by the choice of the stellar yields.
It is known that the yields by TNH give
less iron (about 1/3 for massive stars) and slightly more Mg than the set computed by
WW95, thus providing higher Mg/Fe ratios at a given metallicity
(see also Thomas et al., 1999; Ferreras $\&$ Silk, 2002).
We run our models also with WW95 yields and this implementation
translates in the need of a shorter $\tau$ than those
presented in Table 2 in order not to decrease the mean stellar [Mg/Fe]. 
When trying to evaluate the impact of the yields on the predicted line-strength
indices, the situation is complicated by the adopted calibration.
In particular, for the calibration based on Worthey's (1994) SSPs
a shortening in $\tau$ would not necessarily imply a decrease in the central
$Mg_2$ and $<Fe>$, since they are functions of the Fe abundance alone, which
increases faster when WW95 yields are adopted.

In order to have a more quantitative picture, we show in Table 4 the indices and the mean stellar [Mg/Fe] ratios
predicted by our best model when different yields compilations are chosen. 
In column 1 the model is indicated; in columns 2 and 3 we present the adopted yields compilation and the 
assumed infall timescale, respectively, whereas
in column 4 we show the time at which the galactic wind occurs. In columns 5 to 8 we show
the indices predicted by means of either the Tantalo et al. (T) or the Worthey SSP (W); the last
column shows to the predicted stellar $[<Mg/Fe>]$.
As anticipated, the compilation of yields by WW95
translates in the need of very short infall timescale, in order to match 
the observed Mg overabundance. 
In particular, we have to increase $\nu$ from 10 Gyr$^{-1}$ to 20 Gyr$^{-1}$ or more,
in order to obtain $[<Mg/Fe>] > 0.1$. This result is reinforced by the last row of Table 4,
where we show the predictions based on Pipino et al.'s (2002) best model. This particular model
included WW95 yields and
did not assume any initial gas flow and predicted reasonable values for the central 
metallicity indices
as well as for the the mean stellar [Mg/Fe]. 

On the other hand,
it is worth noting that the use of WW95m set of yields gives the same results (within
a few percent) obtained with our Model II with TNH96. These findings are also
confirmed for the $10^{10}M_{\odot}$ and $10^{12}M_{\odot}$ galaxies, as one can
see from Figure 6 (dashed and long-dashed curves for WW95 and WW95m cases, respectively).
In conclusion, we underline the need for an initial infall episode, which allows us not only to 
build more realistic models, but also to match simultaneously the metallicity indices and the
Mg overabundance in the core of bright ellipticals. As one can deduce from Table
6, our predictions for the central $Mg_2$ and $<Fe>$ are robust and independent
from the adopted yields compilation. The predictions for [Mg/Fe] based on the
best set of yields available
are in reasonable agreement with the observed ones only when the infall term is
present in the equation for the chemical evolution. 

\begin{table*}
\centering
\begin{minipage}{120mm}
\scriptsize
\begin{flushleft}
\caption[]{Model II with different yields for massive stars}
\begin{tabular}{ll|lllllll}
\hline
\hline
Model &Yields & $\tau$& $t_{gw}$ & $Mg_2$ &        & $<Fe>$ &        &$[<Mg/Fe>]$\\
       &      & (Gyr)& (Gyr)     &  T     & W      & T      & W     &          \\
\hline
 II  &  TNH96    & 0.4   &  0.709  & 0.298&  0.285 & 3.33   & 3.17	& 0.206  \\
\hline
 II  &  WW95     & 0.4   &  0.704  & 0.240  & 0.295 & 3.34   & 3.27	& -0.08  \\
 II  &  WW95     & 0.1   &  0.564  & 0.264  & 0.289  & 3.32   & 3.21	&0.04  \\
\hline
 II  &  WW95m    & 0.4   &  0.705  & 0.299  & 0.288  & 3.36   & 3.21	& 0.194\\
\hline
 P2002&  WW95    & -     &  0.439  & 0.299  & 0.274  & 3.26   & 3.07    & 0.253 \\  
\hline
\end{tabular}
\end{flushleft}
The values presented in this table are referred to the central ($0\rightarrow 0.1 R_{eff}$) zone for
a $10^{11}M_{\odot}$ galaxy. 
The indices obtained with the Tantalo et al. (1998) SSP are labeled with T, whereas W refers to the Worthey (1994) SSP (see text). 
TNH96: yields by Thielemann et al. (1996). WW95: yields by Woosley $\&$ Weaver (1995). 
WW95m: yields by Woosley $\&$ Weaver (1995) modified according to Fran\c cois et al. (2003).
P2002: results from Pipino et al. (2002) best model.
\end{minipage}
\end{table*}

In concluding this section we want to stress that
the diagrams presented in Figs. 4 and 6 represent powerful and independent 
tests for our models.
In fact, in order to have high central indices, we need 
a long star formation period to obtain a suitable chemical enrichment; on the contrary, high
values of [Mg/Fe] require very short star formation timescales. To give an example, to produce 
a better agreement between the predicted curve and the observations in Fig. 6, we 
should increase $\nu$ (or decrease $\tau$), but this produces lower values for $Mg_2\, \rm and\, <Fe>$.

\subsection{Metallicity gradients inside the galaxies}
Once we have defined the influence of the total initial mass of the objects in determining
the parameters which regulate galactic evolution, we intend to investigate the variation 
of the star formation and gas accretion histories among different
regions of the same galaxy. 
In order to extract information on the mechanism of galaxy formation,
we make use of the radial gradients in line-strength indices and colors as
measured in ellipticals. 
In this section we focus on the variation of metallicity as a function of the galactocentric distance,
whereas in section 3.4 we study the color gradients.
 
Metallicity gradients are observed in the stellar populations 
inside ellipticals (e.g. Carollo et al., 1993; Davies et al., 1993; Trager et al., 2000a). 
They can arise as a consequence of a
more prolonged star formation, and thus stronger chemical enrichment, in the inner zones. In the 
galactic core, in fact, the potential well is deeper 
and the wind develops later relative to the most external regions (e.g. Martinelli et al., 1998).  
From an observational point of view, however, the situation is not very clear. In fact,
Davies et al. (1993) did not find any correlation linking the gradients to the mass 
or the $(Mg_2)_0$ of the galaxies, whereas Carollo et al. (1993) claimed a bimodal trend with mass,
in which the $Mg_2$ gradient becomes flatter in high mass ellipticals. On the other hand,
Gonzalez $\&$ Gorgas (1996) found that the gradient correlates better with $(Mg_2)_0$
than with any other global parameter. Recently, Kobayashi $\&$ Arimoto (1999) analysed
a compilation of data in the literature, finding that the metallicity gradients do not correlate
with any physical property of the galaxies, including central indices and velocity dispersion, as well
as mass and B magnitude.

In what follows we present only the results for Model II, which seems to be the best 
in reproducing the mass-metallicity relation and the Mg enhancement, 
and we try to discriminate among the different sub-models (for their definition
see section 2.5 and Table 5).
In particular, we study a galaxy of $10^{11}M_{\odot}$ luminous mass by means of Model IIb, which assumes a $\tau$ decreasing from the 0.4 Gyr in the inner zone,
to 0.01 Gyr at $R=R_{eff}$, whereas in Model IIc we adopted $\tau$= 0.5 Gyr at the effective radius.
A summary of the sub-model properties for three particular shells (in order to sample the core, the
$R\simeq R_{eff}/2$ zone and the $R\simeq R_{eff}$ region, respectively) of a $10^{11}M_{\odot}$ galaxy,
is given in Table 5. In column 1 both the internal and the external radius defining that
particular shell are shown; in column 2 we present the infall timescale at that radius, whereas
in column 3 we show the time at which the galactic wind occurs. In columns 4 to 7 we show
the indices predicted by means of either the Tantalo et al. SSP (T) or the Worthey SSP (W); the last three
columns present the predicted mean stellar [$<Mg/Fe>$], [$<Mg/H>$] and [$<Fe/H>$],
respectively.
However, conclusions similar to the results presented here can
be drawn for galaxies in the whole mass range.
We recall that the central zone is the same in all the
sub-models, since it is constrained by the results on central indices and [Mg/Fe].

\begin{table*}
\centering
\begin{minipage}{120mm}
\scriptsize
\begin{flushleft}
\caption[]{Model II - Summary of sub-model properties}
\begin{tabular}{l|ll|llll|lll}
\hline
\hline
Sub-model/	& $\tau$& $t_{gw}$ & $Mg_2$       & $Mg_2$       &	$<Fe>$	& $<Fe>$ & [Mg/Fe] & [Mg/H] &[Fe/H]\\
zone    	& (Gyr) & (Gyr)    &T             &   W          &   T           &  W    &       \\
\hline
a & & & & & & & \\
\hline
$0-0.1 R_{eff}$& 0.4   &  0.709  & 0.298  &0.285  & 3.33  &3.17  &0.206 & 0.07 & 0.116\\  
$0.4-0.5 R_{eff}$&0.4  &  0.579  & 0.287  &0.264  & 3.17  &2.98   &0.245 &0.03 &0.007 \\  
$0.9-1 R_{eff}$&0.4    &  0.544  & 0.284  &0.258  & 3.12  & 2.92  &0.258 & 0.02 & -0.03\\  
\hline
b & & & & & & & \\
\hline
$0-0.1 R_{eff}$& 0.4   &  0.709  & 0.298  & 0.285 &3.33   & 3.17  &0.206 & 0.07 & 0.116 \\  
$0.4-0.5 R_{eff}$& 0.2   &0.490  &0.293   & 0.257 &3.13   & 2.92  &0.306 & 0.03 & -0.03  \\  
$0.9-1 R_{eff}$&0.01   & 0.308   & 0.317  & 0.236 &3.04   & 2.72  & 0.532& 0.02 &-0.145   \\      
\hline
c & & & & & & & \\
\hline
$0-0.1 R_{eff}$& 0.4   &  0.709  & 0.298  &0.285  &3.33   &3.17   &0.206 & 0.07 & 0.116\\  
$0.4-0.5 R_{eff}$& 0.45  & 0.594 &0.287   &0.266  &3.18   &2.99   & 0.236&0.03  & 0.014  \\  
$0.9-1 R_{eff}$&0.5    & 0.569   & 0.283  &0.260  &3.13  & 2.94   & 0.242 & 0.02& -0.016 \\      
\hline
\end{tabular}
\end{flushleft}
Assumed infall timescale and predicted chemical abundances and line-strength indices
as a function of radius for a $10^{11}M_{\odot}$ Model II galaxy.
The calibrations used for $Mg_2$ and $<Fe>$ marked with T are derived from the Tantalo et al. (1998) SSPs, 
whereas those labeled with W are based on Worthey's (1994) SSPs, see section 3. The abundance ratios in the last three
columns are averaged over the stellar populations.
\end{minipage}
\end{table*}

\begin{figure}
\epsfig{file=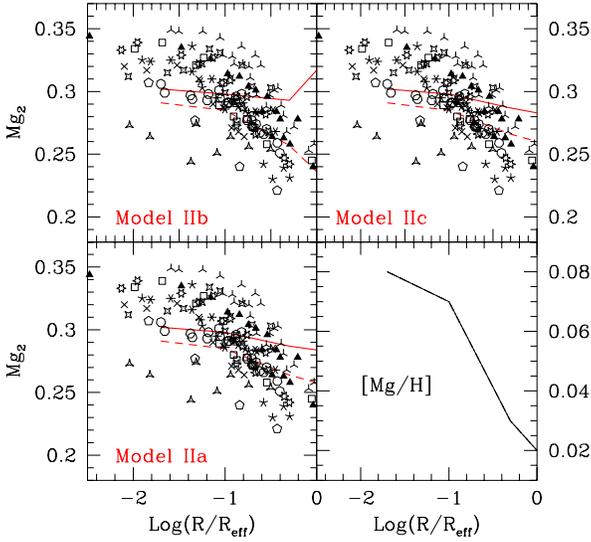,height=8cm,width=8cm}
\caption{$Mg_2$ gradient predicted by different sub-models 
for a $\sim 10^{11}M_{\odot}$
galaxy compared with observations. Each symbol correspond to one galaxy from Carollo et al. (1993). 
Predictions for both Worthey (dashed line) and Tantalo et al. (solid line) calibration
are plotted. The typical error for $Mg_2$ is $\sim 0.002 mag$ in the innermost zone and  $\sim 0.01 mag$ 
at larger radii. In the bottom-right panel, the predicted [Mg/H] abundance gradient is shown for the
three sub-models (see text for details) and the same galactic mass.\label{fig4}}
\end{figure}

We show our results for the predicted trend of $Mg_2$ as a function of radius in Fig. 7 for a $10^{11}M_{\odot}$ galaxy by means of  Model IIa,b,c. 
All the observed galaxies, shown for comparison, 
have a measured mass of $\sim 10^{11}M_{\odot}$.
Each sub-model is plotted in a different panel. In any case the curves obtained with the 
Worthey calibration seem to agree better with the observed slope.  
In the bottom-right panel of Fig. 7, we show the predicted [Mg/H] abundance gradient as
a function of radius for the different sub-models.
It is worth noting (as evident also from the entries in Table 5) that the 
stellar $[<Mg/H>]$ as a function of radius behaves in the same way
in all the three sub-models, thus producing an identical gradient (${\Delta [Mg/H]\over \Delta \log r} = -0.06$) in all cases. 
On the other hand, 
the predicted $Mg_2$ versus radius appears different according to the assumed calibration (e.g. Worthey or Tantalo et al.), 
especially at large radii; the reason for this relies in the fact that the calibration relations introduce a dependence on the
[Mg/Fe] and [Fe/H] (see Fig. 8) . 
As it can be seen from Fig. 7, we obtain a reasonable fit to the data
only with Model IIb and the Worthey calibration. In general our curves
lie around the average value of $Mg_2$ at a given radius, 
but, from a visual inspection of Fig. 7, it seems that every model predicts flatter slopes
than those deduced from the observational points, especially in the inner zones. 
However $Mg_2$ gradients exhibit high dispersion in their slopes
and central values; therefore, in order to have more robust estimates,
we face this issue in a statistical way at the end of this section.

The fast decrease of the [Fe/H] ratio with radius in Model IIb galaxy (see Table 5
and upper panel of Fig. 8) produces a strong Mg
enhancement in the outer regions, which is responsible for the upturn
in the predicted $Mg_2$ at $R\sim R_{eff}$ (solid curve in the
top-left panel of Fig. 7) when the Tantalo et al. calibration is
adopted.  The reason for this unexpected behaviour, completely at variance with
the observed trend, could reside in the adopted
ranges for [Fe/H] and [Mg/Fe] in Matteucci et al. (1998), who provided a
calibration law by adopting the SSPs of Tantalo et al. (1998a) for specific values of the [Mg/Fe] ratio. 
In particular,
these SSPs are limited to [Mg/Fe]=+0.3 whereas our $10^{11}M_{\odot}$
Model IIb galaxy has [Mg/Fe]=+0.532 in its outermost zone.  On the
other hand, since Model IIb with Worthey calibration produces the best
gradient among the whole set of sub-models, in the following we refer
to this model as to the best one.

This choice is confirmed by the analysis of the $<Fe>$ gradient (Fig. 8). The slope 
predicted by different sub-models is flatter at larger radii, when considering the
Tantalo et al. calibration (lines labeled with T). Also in this case, Model IIb with Worthey's 
relation agrees better with the data. We note, however, that the results obtained with the different 
sub-models and the two calibrations are consistent with each other within $2-3\sigma$ (the typical 
uncertainty is shown on the left side of the Figure 8). In fact, observations
show a large scatter and also
a large range in slopes. In any case, our best model well reproduces
the mean values of the observed gradients. In fact, Davies et al. (1993) mean slope for $<Fe>$ gradient is 
$-0.38 \pm 0.26$, consistent with our predictions for a $\sim 10^{11}M_{\odot}$ Model IIb:
- 0.29 and - 0.45 for the Tantalo et al. and Worthey calibrations, respectively.
It is very important to compare these values with the real abundance gradients.
From Table 5 we derive 
${\Delta [Fe/H]\over \Delta \log r} = -0.119\, , -0.261\, , -0.132 $ for Model IIa, IIb and IIc,
respectively. 
Model IIb shows a very steep gradient in Fe (see small panel of Figure 8), in good agreement with the mean metallicity gradient
found by Kobayashi $\&$ Arimoto (1999) ${\Delta [Fe/H]\over \Delta \log r} = -0.30\pm 0.12$,
and the results of Davies et al. (1993). A fundamental conclusion which can be drawn from the 
discussion presented above, is that, while the variations in $Mg_2$
depend on the radial gradients in [Fe/H] and [Mg/Fe] (see Fig. 8), the predicted
$<Fe>$ gradient reflects instead the variation of the real abundance, [Fe/H], as a function of galactic radius (see Fig. 8).

\begin{figure}
\epsfig{file=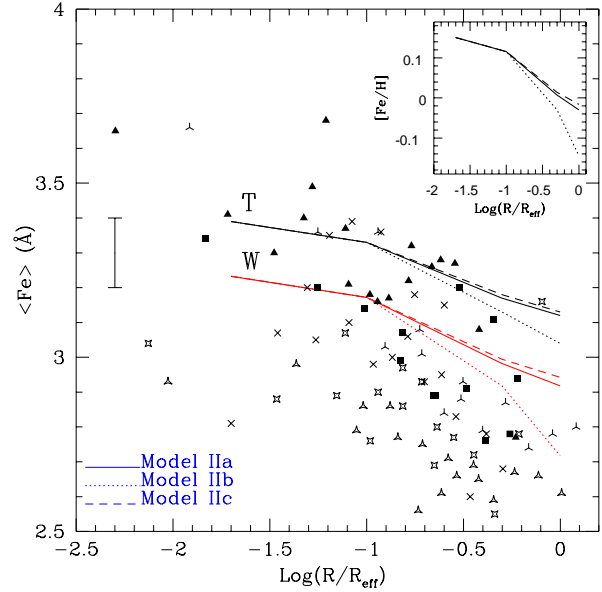,height=8cm,width=8cm}
\caption{$<Fe>$ gradient predicted by different sub-models for a $\sim 10^{11}M_{\odot}$
galaxy compared with observations. Each symbol corresponds to one galaxy from Davies et al. (1993) data.
The predictions based on Worthey calibration are labeled with \emph{W},
whereas those from Tantalo et al. with \emph{T}. The typical error
is shown on the left.
In the small panel in the top-right corner we show the predicted [Fe/H] as a function of radius for the same models.\label{fig5}}
\end{figure}

As a further step in our analysis, we followed the suggestion by Kobayashi $\&$ Arimoto (1999).
In particular, we derived the
slope and the intercepts of line-strength gradients via a least-square fit:
\begin{equation}
Mg_2 (r_i)=(Mg_2)_{eff}+{\Delta Mg_2\over \Delta \log r} \log {r_i\over r_{eff}} \; ,
\end{equation}
where $Mg_2 (r_i)$ is the index predicted in the $i-th$ shell of radius $r_i$ and
the subscript \emph{eff} means that the values are computed at the effective radius $R_{eff}$.
The same procedure was carried out by Kobayashi $\&$ Arimoto (1999) for the observed gradients.
To be consistent with the gradients obtained from the observations we limit our analysis to 0.1 $R_{eff}$,
since Kobayashi $\&$ Arimoto (1999) neglected data taken at inner radii to avoid seeing effects. We applied this procedure
to galactic models in the $10^{10}-10^{12}M_{\odot}$ mass range and our results are plotted in Fig. 9.
With this kind of diagnostic we can check in a statistical way whether the predicted slopes match average value of the observed ones
as a function of the galactic mass, in order to obtain a more quantitative comparison between our Model IIb
and the observed gradients.

\begin{figure}
\epsfig{file=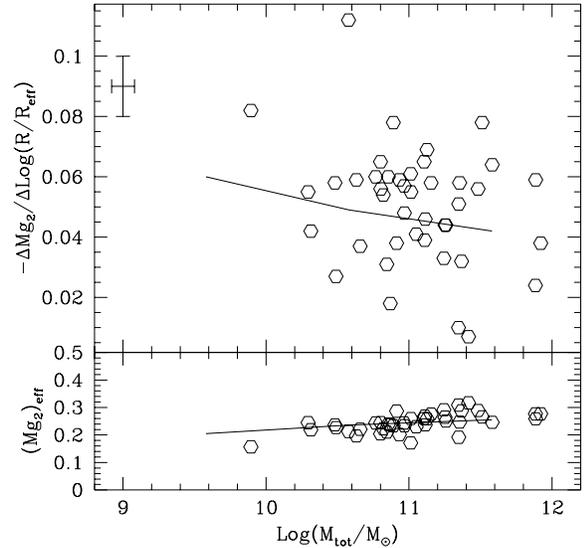,height=8cm,width=8cm}
\caption{$Mg_2$ gradient slope and intercept defined as in eq. 10 from Model IIb. 
Data from Kobayashi $\&$ Arimoto (1999).
\label{fig6}}
\end{figure}

We tuned $\tau$ in order to obtain reasonable values for the indices in each zone.
In particular, we fixed $\tau=0.01\, \rm Gyr$ at $R=R_{eff}$ for the models with initial masses of $\sim 10^{10}M_{\odot}$
(as found for the $\sim 10^{11}M_{\odot}$ case), and  $\tau=0.06\,\rm Gyr$ for the most massive galaxies.
From the bottom panel of Fig. 9 we note the excellent agreement with the average value of $Mg_2$
measured at the effective radius. Our model predicts a logarithmic slope for this index which
is very close to the average observed gradient, although we obtain slightly flatter gradients
for the most massive galaxies. This is a clear consequence of the adopted inverse wind scenario,
in which the more massive galaxies form faster. In any case, this approach 
allows us to show the goodness of our models in a clearer way than that presented in Fig. 7.
The little disagreement with observations can be explained as a consequence of 
the predicted metallicity in the core of the most massive galaxies, which seems
to be smaller than the observed one (at least for what can be deduced
from the comparison made in Fig. 4).
As evident from Figure 9, there is a large scatter in the observed slopes, but on the average  ${\Delta Mg_2\over \Delta \log r}$
seems to be independent from the mass of the galaxies.

As in the case of the central indices, the combined analysis of $Mg_2$ and 
$<Fe>$ variations along the galactic radius can provide a measure of the [Mg/Fe] gradient.
A reliable extimate of the slope of the [Mg/Fe] gradient represents
a very useful tool to discriminate between outside-in and inside-out formation
scenarios. Our best model predicts ${\Delta [Mg/Fe]\over \Delta \log r} = 0.326$ (in the case 
of a $\sim 10^{11}M_{\odot}$ galaxy), whereas model IIa and IIc seem to suggest flatter
slopes. Hints for a positive slope in the most massive galaxies, in agreement with our predictions, 
can be found in the analysis made by Ferreras \& Silk (2002) based on Trager et al. (2000a) data;
however the scatter increases and the mean slope for a given galactic mass decreases
if the Kobayashi \& Arimoto (1999) compilation is added to the Trager et al. (2000a) points. On the other hand,
in a very recent study of Coma early-type galaxies, Mehlert et al. (2003)
claimed that, on average, these galaxies do not exhibit gradients in $\alpha/Fe$,
thus disfavoring models with strong inside-out or outside-in formation
scenarios. Given the importance of the topic, larger compilations of homogeneus
data are required before considering this issue completely solved.

\subsection{Color-Magnitude relation}

In the following sections we focus on the photometric results, in order 
to show
how the good agreement achieved between Model IIb and the abundance patterns
translates into a fit of the observed colors. 
In order to do that, we followed the standard route in the context of the 
monolithic collapse, 
namely the CMR arises directly from the mass-metallicity relation
(i.e. stellar populations with higher metal content show redder colors).
Moreover, the very small scatter observed in the colors as a function of the total V magnitude
for galaxies in Virgo and Coma clusters (Bower et al., 1992),
and the fact that it is observed in different environments and at different redshift (e.g. Stanford et al., 1998;
Bernardi et al., 2003)
are interpreted as evidences of an highly syncronized epoch of elliptical formation
at high redshift (Bower et al., 1992; Renzini $\&$ Cimatti, 1999). 
However, from a theoretical point of view, the well-known age-metallicity degeneracy implies that 
the presence of a significant fraction of younger (i.e. bluer) stellar populations 
can be masked if they have a high metal content. This kind of age-metallicity conspiracy
has been claimed by several authors (e.g. Trager et al., 1998, 2000; 
Ferreras et al., 1999; Jorgensen, 1999, but see Kuntschner, 2000, and Kuntschner et al., 2001,
for a different point of view) in order to reconcile the tightness of the CMR
with evolutionary model in which the star formation continues until low redshifts,
as a consequence of late mergers. The question is still under debate
and we intend to assess this issue separately in a forthcoming paper. 
Therefore, we assume our galaxies to form simultaneously at 
high redshift. We recall that the epoch of galaxy formation adopted in this paper is $z_f =5$ 
(i.e. our model ellipticals are 12.3 Gyr old).

In a future paper we shall analyse galactic models in which alternative scenarios
of galaxy formation will be assumed. In particular we intend to test 
the possible role of merging, late episodes of gas accretion, secondary bursts of star formation in
the CMR and the chemistry of ellipticals. 

\begin{figure}
\epsfig{file=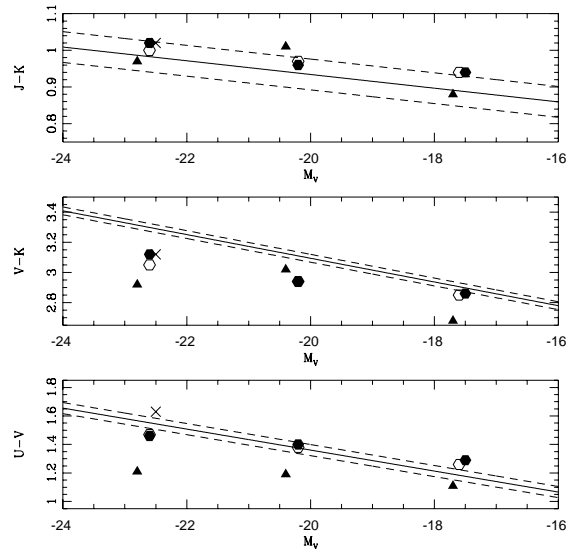,height=8cm,width=8cm}
\caption{Predicted CMRs for Model I (empty hexagons), Model II (full hexagons) and the best model
by Pipino et al. (2002, triangles). The solid lines represents the fits of Bower et al. (1992b) data,
with their 1$\sigma$ scatter (dashed lines). 
The prediction
from the model with fixed metallicity (Z=0.03, see text) is shown by a cross.\label{fig8}}
\end{figure}
Before presenting the photometric results, we recall that Jimenez et al. (1998) 
photometric code was applied to the SSPs predicted by our models, thus taking into account 
in a self-consistent way the chemical evolution of the ISM.

Fig. 10 shows the predicted Color-Magnitude relations for U-V, V-K and J-K colors (Model I and 
and II) versus total
V magnitude compared to the observed data which are represented by their best fit (Bower et al., 1992b; solid lines).
The predicted colors are obtained by adding the different shells contributions 
up to $\sim$5 kpc from the galactic center, whereas the total $M_V$ is relative to
the whole galaxy (up to 10 $R_{eff}$), in order to be consistent with
Bower et al. (1992a), who obtained their data on colors with a fixed aperture of 5$h^{-1}$ kpc,
whereas, for the V magnitude, they used its \emph{total} value
(see Kawata, 2001, for a detailed discussion). This procedure 
becomes necessary for a sampling of the central regions in the high mass 
models, since it simulates
the observed aperture and it can be done simply by
avoiding the inclusion of the most external parts, where
the presence of the color gradients would make the predicted colors bluer.
Model I and II agree very well
with the data in the J-K and U-V versus $M_V$ diagrams. We note a slightly 
flatter trend for our curves relative to the data, but still consistent 
within the scatter ($\sim 0.04$ mag, Bower et al. 1992b, dashed lines in Fig. 10) 
of the observed galaxies around the mean relation. We recall that also
the slopes obtained from the fit to the data have their own error of 0.01 (see Bower et al. 1992b).
For comparison, we plot also the CMR predicted by  Pipino et al. (2002) best model
(triangles). In spite of the fact that this model is very similar to the ones presented in this paper, 
it does not consider infall, thus it produces a larger fraction of stars formed from  
low-metallicity gas than our Model II. Consequently, the lower mean metallicity
predicted for the stellar populations in the best model of Pipino et al. (2002) makes the galactic light bluer.
We do not show Model III results, since they fail in reproducing the CMR (e.g.
$V-K\sim 3.7$ for the more massive galaxies) due to a too high mean stellar metallicity.

\begin{figure}
\epsfig{file=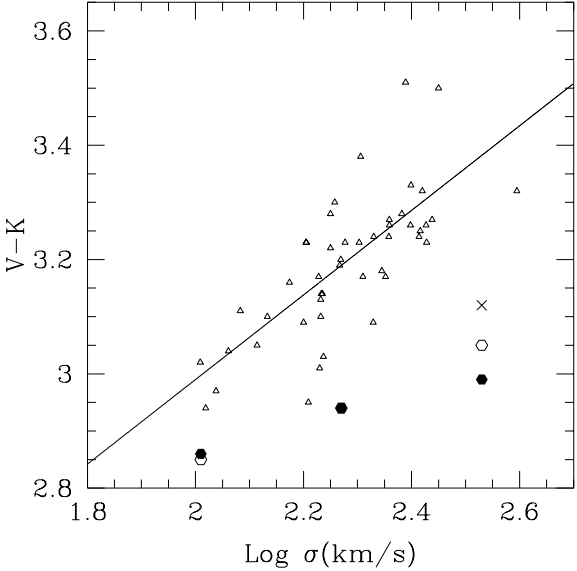,height=8cm,width=8cm}
\caption{Predicted V-K versus velocity dispersion for Model I (empty hexagons) and II (full hexagons)
compared to the data from Mobasher et al. (1999, triangles). The line represent the fit of the data. The prediction
from a model with fixed metallicity (Z=0.03, see text) is shown by a cross.\label{fig9}}
\end{figure}

On the other hand, our model predictions fail in reproducing the right slope
for V-K versus $M_V$. In particular, 
we found a V-K which is about 0.5 mag less than the observed one for the most massive galaxies. 
This problem becomes more evident in Fig. 11, where we plot the color as a function of the galactic velocity dispersion. 
However, we consider this particular plot less constraining than the color-$M_V$ diagram, since
the predicted relation could be sistematically biased by the relation adopted to convert  
the total magnitude of the object into the observed $\sigma$. 
Several authors in the literature (e.g Arimoto $\&$ Yoshii, 1987; Romano et al., 2002)
pointed out that, 
since the V-K color has a strong dependence on the global metallicities, 
models with Salpeter IMF hardly match the observed CMR, producing in general bluer galaxies.
On the other hand, the fact that the V - K predicted by the best model of Pipino et al. (2002,
with inverse wind but without infall) flattens at high luminosities, reinforces the conclusion
that the disagreement is caused by a too low predicted metallicity for the cores
of the most massive ellipticals with respect to observations, 
and not by the presence of the initial infall episode.
We tried to reconcile our predictions with the data
by varying \emph{ad hoc} the mean stellar metallicity ($<Z>$), while we kept fixed the IMF and the
star formation history for the $10^{12}M_{\odot}$ Model II galaxy. In Table 6 we show 
the colors obtained with several fixed metallicities compared to our case 
(Model IIb) which predicts
$<Z>=0.025$. 
Since the differences in the stellar evolution prescriptions included in the photometric codes 
could generate 0.25 mag discrepancies in V-K (Charlot et al., 1996), we checked our finding by comparing it with
the results obtained with the Bruzual $\&$ Charlot (1993) code with solar metallicity,
applied to our Model II.
As it can be seen from the last row of Table 6, also in this case the predictions
do not agree with the observed values.
We reject the high metallicity models, which predict redder colors 
than the typical values of the observed CMR (see the entries in Table 6) for a given $M_V$. 
On the other hand, the $<Z>=0.02$ model predictions
are very similar to our 'self-consistent' case, as expected, since they have nearly the same 
metallicity. The model with $<Z>=0.03$ (cross in Figs. 10 and 11) gives a better fit
for the V-K - $M_V$ relation, but makes the agreement slightly worse for the other two colors.
In conclusion, in order to have a good fit to the three CMRs, it is necessary to
increase the mean stellar metallicity from $<Z>=0.025$ to $<Z>=0.035$ at most. 
This can be obtained by assuming a slightly flatter IMF with exponent x=1.25, which gives
$\rm Mg_2 =0.312\, mag $, $<Fe> = 3.42 \rm\, \AA$ and $\rm [Mg/Fe]=0.341\, dex$ in the central zone
(still consistent with observations) and  $<Z>=0.037$ within the inner 5 kpc.

\begin{table}
\scriptsize
\begin{flushleft}
\caption[]{The effect of metallicity on colors.}
\begin{tabular}{l|lllllll}
\hline
\hline
$<Z>$ & $M_V$ & $L_B$              &$L_K$               & U-V & V-K & J-K\\
      &       &($10^{10}L_{\odot}$)&($10^{10}L_{\odot}$)&     &     &     \\
\hline
0.02  & -22.7 &7.78                &40.5                &1.49 &3.03 & 0.99\\
0.03  & -22.5 &6.29                &41.2                &1.63 &3.12 & 1.02\\
0.04  & -22.3 &4.80                &41.9                &1.76 &3.39 & 1.10\\
0.10  &-21.8  &2.93                &47.6                &2.30 &3.34 & 1.26\\
\hline
Model II&-22.6&	7.42	           & 40.6                &1.46 &3.04 & 1.00\\
\hline
Model BC&-22.2&	5.49	           & 26.9                &1.41 &3.02 & 0.87\\
\hline
\end{tabular}
\end{flushleft}
Luminosities and colors predicted by a $10^{12}M_{\odot}$ galaxy of Model II, compared
to models with fixed metallicity, but same star formation history. Model BC: Model II
colors predicted by means of Bruzual $\&$ Charlot (1993) code with solar metallicity.
\end{table}

\subsection{Color gradients}

The analysis of the color gradients in the stellar populations of elliptical galaxies 
can reinforce the results found in the previous sections.
It is, in fact, observed (e.g. Peletier et al. 1990)
that stars in the galaxy centers are redder than those in the outer
regions, and colors become progressively bluer with increasing radius.
Also in this case, the variation of the gradient strength as a function of redshift
seems to be consistent with passive evolution (Saglia et al., 2000).
The analysis of these gradients out to redshift $\sim 1$ led Tamura et al. (2000) to the
conclusion that they originate mainly from the variations
in the mean stellar metallicity rather than from an age sequence. 
Following the same procedure
adopted in section 3.2 to compare the predicted metallicity gradients with observations,
we show in Figs. 12 and 13 the slopes and the intercepts at $R=R_{eff}/2$ for U-R and B-R color gradients
as predicted by Model IIb and compared with the data from Peletier et al. (1990).

\begin{figure}
\epsfig{file=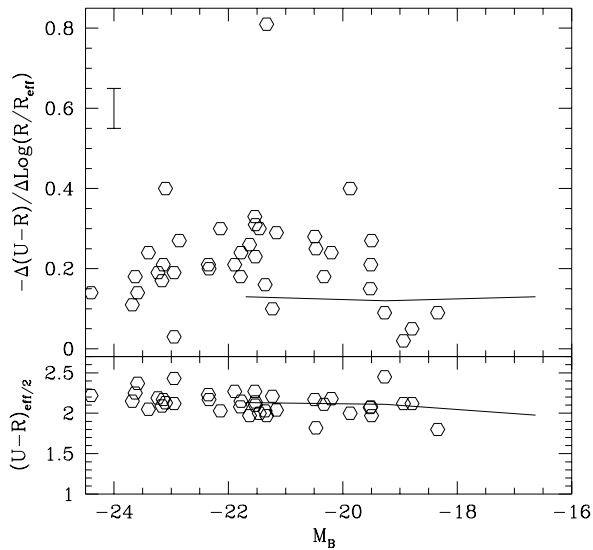,height=8cm,width=8cm}
\caption{Predicted U-R color gradient as a function of galactic total $M_B$ for Model IIb compared with the data
from Peletier et al. (1990). \label{fig10}}
\end{figure}

\begin{figure}
\epsfig{file=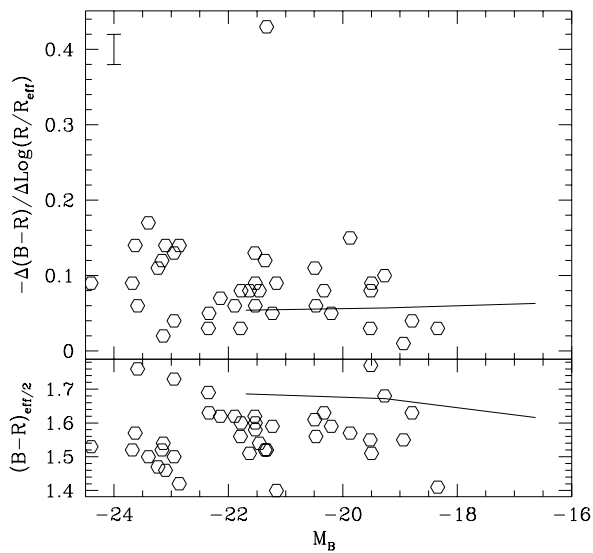,height=8cm,width=8cm}
\caption{Predicted B-R color gradient as a function of galactic total $M_B$ for Model IIb compared with the data
from Peletier et al. (1990). \label{fig11}}
\end{figure}

As expected from the good agreement between our models and the line-strength gradients, 
we note from the figures that the predicted intercepts at $R=R_{eff}/2$, as well as the slopes, 
are pretty close to the average value of the observed ones.
With the help of this diagnostic we can rule out the sub-model c with $\tau<0.2\,\rm Gyr$ 
in the inner zone, since it produces positive color gradients due to the presence of
galactic winds which start earlier in the core than in the outskirts of a galaxy.
Therefore, we can rule out inside-out mechanisms for the formation of elliptical galaxies similar
to those invoked for spiral discs (e.g. Matteucci \& Fran\c cois, 1989;
Chiappini et al., 1997), in which a strong (i.e. $\Delta \tau >>1\,\rm Gyr$) variation 
in $\tau$ between the innermost and the outermost
regions is assumed.
In this case, the timescale for the infall is increasing with galactic radius,
inducing the external regions to evolve much more slowly than the internal ones.
A large variation in age between the inner and the outer stellar populations
seems to be ruled out also by the recent results of Mehlert et al. (2003, see also Sec 3.2).

It is worth noting that
Menanteau et al. (2001a) showed that a fraction of field early-type galaxies exhibit
blue cores. The properties of these galaxies can be
reproduced by means of Martinelli et al. (1998) multizone model,
which is an outside-in formation model as the one presented here,
and by
assuming that only $\sim 20\%$ of the galaxies have been formed very recently (Menanteau et al., 2001b), being the
blue color of the central region of the galaxies associated to more prolonged star formation in the core with respect to 
the outskirts.
Moreover, it is well known that field ellipticals as a whole are $\sim 1-2$ Gyr 
younger than the counterpart living in clusters (e.g. Bernardi et al., 1998). The
presence of a late episode of star formation in the galactic core can be explained
in terms of accretion/merging events or by a residual star formation which survived
the development of the galactic wind in the very center of the galaxy.

\subsection{Mass to light ratio}

In this last section we deal with the fundamental plane and,
in particular, we focus on the possible variation of the mass-to-light ratio as a function
of the mass of the galaxy (FP seen e\emph{dge-on}). We recall that
if ellipticals were a perfect homologous family, the FP could be explained
by means of the virial theorem (e.g. Renzini $\&$ Ciotti, 1983). The difference from the expected relation in the plane
seen edge-on is called the \emph{tilt} of the FP. The FP edge-on is
simply the projection of the locus in which ellipticals lie on the plane
formed by the two parameters $\kappa_1$ and $\kappa_3$ linked to the mass and to the mass-to-light ratio (M/L),
respectively (see Bender et al., 1992; Burstein et al., 1997). Therefore the observed tilt
translates into a systematic
variation of the M/L with the mass or the luminosity.

\begin{figure}
\epsfig{file=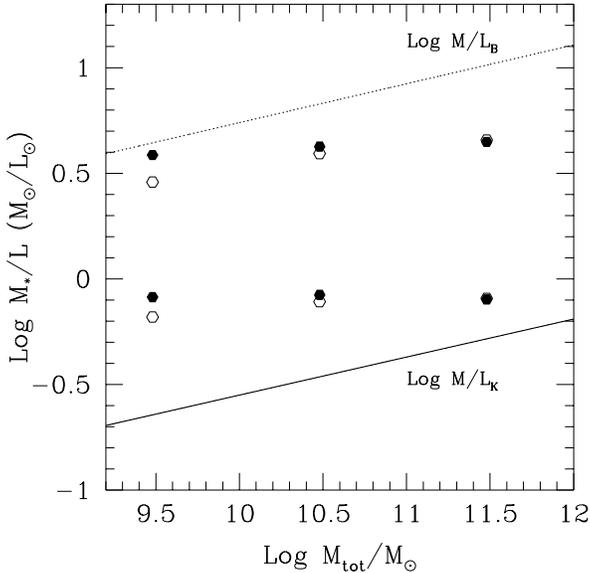,height=8cm,width=8cm}
\caption{Predicted M/L in B- and K-band for Model I (empty hexagons) and Model II (full hexagons)
as a function of the galactic mass. Solid line: fit from Mobasher et al. (1997) data in the near-infrared.
Dashed line: fit from Burstein et al. (1997) for the B-band.\label{fig12}}
\end{figure}

Our results in the B and K-band are plotted in Fig. 14, where the solid line represents the 
fit to Mobasher et al. (1997) data in the near-infrared, and the
dashed line the fit to Burstein et al. (1997) data in the B-band derived at the effective radius. 
The predicted luminosities
for our galaxies are given in Tables 1 and 2, whereas the \emph{stellar} masses adopted in deriving the mass-to-light
ratio are $\sim 20-30 \%$ of the nominal mass. In both bands our models exhibit a
flat behaviour at odds with observations, and this is because we assume a constant IMF. 
However the average theoretical $M/L_B$ values are close
to the observed ones, whereas the $M/L_K$ values suffer from the effects of the too low $L_K$
predicted by our models. In fact, since we are using the stellar mass, i.e. neglecting the
contribution of the dark matter, the only solution (at fixed IMF) in order to match the
observed $M/L_K$ is to increase the K-band luminosity. 
In order to solve this discrepancy, one possibility could be a change in the mean stellar metallicity of the same 
factor we need to match the V-K color in high mass ellipticals.
By using the B-luminosities displayed in Table 6, we infer a increase in $log (M/L_B)$
of the order of $\sim 0.1-0.2$ for a $<Z>=0.035$ model. If this change is applied only
to the most massive galaxy, it could help in reproducing the right slope.

A solution often invoked to explain the tilt of the FP is a change 
in the IMF, which has to become flatter
with increasing initial mass (e.g. Renzini $\&$ Ciotti, 1993; 
Matteucci, 1994). In this case we still have a change in metallicity,
but also a decrease in the low-mass stars contributing to the mass 
observed at the present time.
If we consider a case in which the $10^{10}M_{\odot}$
galaxy of Model II has a Salpeter IMF and the $10^{12}M_{\odot}$
galaxy an AY one, we obtain a $M/L_B$ ratio increasing of about a factor of 1.3, with a slope
which becomes steeper and similar to the observed one. On the other hand, the $log (M/L_K)$ 
decreases to the value of -0.32, thus reaching the observed relation, but producing a decreasing
trend in $log (M/L_K)$ as a function of mass, completely at variance with the solid line in Fig. 14.
In any case, such kind of model
would fail in reproducing the CMR (see section 3.3), and the
$Mg_{2}-$mass relation (with $(Mg_{2})_0 > 0.35 mag$  for the most massive galaxies).
We also tried the intermediate IMF used in section 3.3.

However, recent results from Scodeggio et al. (1998) seem to question
the tilt of the FP, at least in the near infrared. After correcting for 
the completeness, 
they found $M/L_B \propto L_B^{0.18}$ and $M/L_K \propto L_K^{0.08}$. Furthermore
they claim zero tilt in the K-band in case of fully corrected samples. In this case
the increasing slope of the mass-to-light ratio for bluer bands is driven by the CMR.
Finally, other possible solutions to the tilt of the FP not investigated in the present paper, invoke e.g. deviation
from homology (e.g. Busarello et al., 1997), or changes in the DM content as a function of luminous mass 
(see Ciotti $\&$ Renzini, 1993).

\section{Discussion and Conclusions}

In this paper we have demonstrated how the majority of the observables in ellipticals
can be reproduced within a reasonable agreement by using a multizone model
for the photo-chemical evolution of elliptical galaxies based on the 
standard collapse of a gas cloud on short timescales
occurring at high redshift.
In particular, we fixed the free parameters in our best model
(namely the collapse timescale $\tau$ and the efficiency of star formation $\nu$), 
by requiring that
they can reproduce the mass-metallicity and the color-magnitude relations, as well
as the Mg enhancement in massive galaxies and the metallicity gradients.
After a detailed analysis, we ruled out models with a flat IMF (x=0.95, Model III) or
with infall timescale constant with galactic mass (Model I). Our best model
has a Salpeter IMF, a star formation efficiency increasing with galactic mass
and a shorter infall timescale for the most massive objects.

Before concluding, we discuss here some of the assumptions made in this paper and 
their consequences for the theory of galaxy formation and evolution.

First of all, we focus on the relative
weights that the star formation efficiency and infall timescale
have in driving the mass-metallicity relation.
In particular, our assumptions on $\tau$ and $\nu$ can be discussed in the light of previous
results. This issue, in fact, has been recently investigated also
by Ferreras $\&$ Silk (2003).
With the help of a simple model of star formation they explore the correlation
between U-V colour and [Mg/Fe] ratios and conclude that 
a scenario where
$\nu \propto M_{lum}$ should be favoured.
Thomas et al. (2002)
favour also a similar model, in which the more massive objects formed stars at very high
redshift, with a typical duration of the star formation of $\sim 0.5$ Gyr, whereas less massive galaxies
could exhibit star formation even at $z<1$. 
Moreover, Ferreras $\&$ Silk (2003) conclude that the effects of $\nu$ and
$\tau$ are degenerate, so they can be at work together, in a scenario in which
$\tau \propto 1/ \sqrt M_{lum}$. 
An infall timescale decreasing with galactic mass can be physically motivated on the basis of the
following arguments. First of all, in a galactic formation scenario in which
the galaxies are assembled starting from gas clouds falling onto the galactic potential well,
the more massive objects have a higher probability (i.e. a higher cross section)
to capture the clouds than the less massive ones, thus completing their assembly in a faster timescale
(Ferreras $\&$ Silk, 2003) . 
This picture is consistent with the other fundamental hypothesis of this paper, namely
that $\nu$ increases with the galactic mass. In fact,
the inverse wind picture can be explained by  
higher densities or higher cloud-cloud collision velocities in more massive ellipticals, determined
by the assembly of these galaxies by means of the merging of gaseous protoclouds (Matteucci, 1994).
Moreover, as a consequence of the more intense
star formation, the more massive galaxies experience a stronger SN feedback with respect to the less massive ones.
Therefore, the gas in the hot phase is able to halt the cold infalling gas earlier
in a $10^{12}M_{\odot}$ model than in a $10^{10}M_{\odot}$ galaxy, thus making shorter the accretion timescale of
the former with respect to the latter.

This is in contrast with the belief
that the infall timescale should be proportional to the free-fall time, as suggested by several authors
(e.g. Tantalo et al. 1998b, Romano et al. 2002). In fact,
Tantalo et al. (1998b) assumed $\tau$ to be an increasing function
of the galactic mass, 
while their $\nu$ increases with decreasing galactic mass and, within
the same model, going towards larger radii. This is necessary, in their model, 
to create metallicity gradients inside the galaxies. In fact, although the physical meaning of this choice is not clear, the increase
in star formation efficiency with radius counter-balances 
the increase in the infall timescale and preserves the metallicity gradient.
We did not explore this possibility, since an increasing $\nu$ with galactocentric radius
would mean that the star formation is more efficient where the gas density is lower, at variance with 
observations and with the concept of the 'inverse wind'.

On the other hand, Romano et al. (2002) avoided this problem 
by assuming that the star formation has a shorter duration in more massive systems due 
to the appearence of a quasar phase, occurring first in the most massive systems.

Finally, in a recent paper, Kawata $\&$ Gibson (2003) supported the conclusion that $\tau$
should increase with the galactic mass 
by means of chemodynamical simulations. They found that radiative cooling
becomes more efficient and thus the gas infall rate increases with decreasing
galactic mass. However, they did not reproduce the [Mg/Fe] increasing with galactic 
mass, since they did not assume $\nu$ increasing with mass.


In conclusion, 
the $\tau - M_{lum}$ relation we found could seem counter-intuitive in explaining the mass-metallicity
relation, particularly if we see the problem under a classic-wind-perspective. Nevertheless, we underline that
despite of the shorter star forming period, more massive models have a more efficient
star formation (in our models it could be about an order of magnitude higher, see Tables 1, 2 and 3)
and, thus, a faster metal production which can produce the correct mass-metallicity relations and CMRs.

Only when we compare our predictions for line-strength indices and colors for the $10^{12}M_{\odot}$ case, we 
obtain a too low predicted mean stellar metallicity. The very short star formation timescale, constrained
by the [Mg/Fe] ratio, does not allow the galaxies to reach high values for the central $Mg_2$, and this could be a limit
of the inverse wind picture; however, we recall that these conclusions may depend on the assumed index calibration
(see section 3.2) and on the adopted yields (see section 3.1.2), once the IMF has been specified.

A possible way of solving the problem of the too low central metallicity predicted in the core of massive ellipticals
is perhaps suggested by our best model. Since we showed that a model with the infall timescale decreasing 
with radius is the best one in reproducing the observed gradients in ellipticals, we can imagine
a scenario in which a galaxy forms very fast in its external parts, while the inner regions are still 
accreting gas and forming stars. In other words, this resembles to the scenario proposed by Larson (1974) for the formation of ellipticals. 
In this scenario, the first stars form all over the original collapsing gas cloud 
and while the stars remain where they formed, the residual gas keeps collapsing towards the center. In this picture, the gas enriched in metals from the first stellar generations accumulates in the galactic center where the star formation lasts longer.
In our model, the star formation is then inhibited first in the outskirts of the galaxies
and later in the inner regions, 
owing to the heating of the ISM and consequent galactic wind.
However, we do not consider the possible radial flows  connected to the collapsing gas which settles 
towards the center. 
This mechanism could reconcile our predictions for the central indices with the
observations and provide a natural explanation for our findings.

Another possibility for obtaining a higher central metallicity, 
could be a slightly flatter IMF in more massive galaxies.
As we infer from the analysis of the CMR (Sec. 3.3) and of the M/L (Sec. 3.5) an exponent
x=1.25 could be the case for a $10^{12}M_{\odot}$ model.

Our main conclusions can be summarized as follows:
\begin{enumerate}
\item Starting from a fast assembly of gaseous clouds at high
redshift, we obtain $\sim 12$ Gyr old galaxies, suffering
a galactic wind at early times ($t_{gw} \le 1.3$ Gyr) and undergoing passive evolution since then,
whose properties match very well many observations. In particular, 
our best model reproduces the Mg and Fe abundances in the cores
as well as the Mg overabundance relative to Fe, as inferred from
the analysis of the line-strenght indices. 
The multi-zone formalism
allows us to predict the variation of these abundances and abundance ratios
with radius. We predict abundance gradients in good agreement with observations,
once a transformation of abundances into indices is made. 
Furthermore we find that the observed Ca underabundance relative to Mg can be real,
due to the non-neglibile contribution of type Ia SN to the
production of this element.

\item In the framework of the 'inverse wind' model we suggest that the most massive ellipticals
formed stars for a shorter period than less massive ones. This mechanism implies that the most massive objects are older than the
less massive ones, in the sense that larger galaxies undergo passive evolution at earlier times.
As a consequence of this and the different roles played by type Ia and II SNe,
we are able to reproduce the observed increase of the 
average stellar [$<Mg/Fe>$] with galactic mass (e.g. central velocity dispersion).

\item The photometric properties such as the color-magnitude relation
and the color gradients have a natural explanation as due to variations
in metallicity.

\item The Salpeter (1955) IMF seems to be the best one
in order to reproduce the mean photo-chemical properties of elliptical galaxies.
We recall here that models with
the Salpeter IMF are also successful in reproducing the metals
and the energy of the intracluster medium (Pipino et al., 2002) as well as the chemical abundances
in Lyman-break galaxies (Matteucci $\&$ Pipino, 2002).
Only in high mass models a slightly flatter IMF seems to be required
to match the observed Color-Magnitude relation. 

\item The initial infall episode seems to be necessary to explain
the observed [Mg/Fe]$\sim 0.1-0.4$ dex in the central cores of elliptical galaxies.
In fact, closed-box models would predict, for the same stellar yields, too high values for the [Mg/Fe] ratio.
Furthermore, since the gas accretion modulates the star formation rate, the galactic wind and 
the ISM enrichment history are strongly influenced by the infall.
Therefore, in our models, the initial accretion episode represents one of the main drivers
of the mass-metallicity relations.

\item Elliptical galaxies have probably formed outside-in, i.e. the most 
external regions are older and stop forming stars
before the central ones, a suggestion which needs to be tested 
with detailed chemo-dynamical models.

\end{enumerate}

\section*{Acknowledgments} 

We are indebted with R. Jimenez for providing his photometric code, and to 
D. Thomas for his data on [Mg/Fe]. We thank the anonymous referee
for his careful reading of the paper.
Finally we wish to thank F. Calura and C. Chiappini
for interesting discussions.

\label{lastpage}


\begin{thebibliography}{}

\small
\bibitem []{}Anders, E., $\&$ Grevesse, N., 1989, Geochim. Cosmochim. Acta, 53, 197
\bibitem []{}Arimoto, N., $\&$ Yoshii, Y. 1987, A$\&$A, 173, 23 (AY)
\bibitem []{}Bekki, K., $\&$ Shioya, Y. 1999, ApJ, 513, 108
\bibitem []{}Bender, R., Burstein, D., $\&$ Faber, S.M. 1992, ApJ, 399, 462
\bibitem []{}Bernardi, M., Renzini, A., da Costa, L.N., Wegner, G., Alonso, M.V.,
Pellegrini, P.S., Rit\'e, C., \& Willmer, C.N.A. 1998, ApJ, 508, L143
\bibitem []{}Bernardi, M., Sheth, R.K., Annis, J., et al. 2003, AJ, 125, 1882
\bibitem []{}Bertin, G., Saglia, R.P., Stiavelli, M., 1992, ApJ, 384, 423 
\bibitem []{}Beuing, J., Bender, R., Mendes de Oliveira, C., Thomas, D., Maraston, C., 2002, A$\&$A, 395, 431
\bibitem []{}Bower, R.G., Lucey, J.R., Ellis, R.S., 1992a, MNRAS, 254, 589
\bibitem []{}Bower, R.G., Lucey, J.R., Ellis, R.S., 1992b, MNRAS, 254, 601
\bibitem []{}Bressan, A., Chiosi, C., Fagotto, F. 1994, ApJs, 94, 63
\bibitem []{}Bruzual, G., $\&$ Charlot, S. 1993, ApJ, 405, 538
\bibitem []{}Butcher, H.R., $\&$ Oemler, A. 1978, ApJ, 219, 18
\bibitem []{}Burstein, D., Bender, R., Faber, S.M., $\&$ Nolthenius, R. 1997, AJ, 114, 1365
\bibitem []{}Busarello, G., Capaccioli, M., Capozziello, S., Longo, G., \& Puddu, E. 1997, A\&A, 320, 415
\bibitem []{}Cappellaro, E., Evans, R., Turatto, M., A$\&$A, 1999, 351, 459
\bibitem []{}Carollo, C.M., Danziger, I.J., $\&$ Buson, L. 1993, MNRAS, 265, 553
\bibitem []{}Cioffi, D.F., McKee, C.F., $\&$ Bertschinger, E., 1988, ApJ, 334, 252
\bibitem []{}Charlot, S., Worthey, G., Bressan, A. 1996, ApJ, 457, 625
\bibitem []{}Chiappini, C., Matteucci, F., Gratton, R. 1997, ApJ, 477, 765
\bibitem []{}Chiosi, C., $\&$ Carraro, G. 2002, MNRAS, 335, 335
\bibitem []{}Colless, M., Burstein, D., Davies., R.L., McMahan, R.K.Jr, Saglia, R.P., Wegner, G. 1999, MNRAS, 303, 813
\bibitem []{}Daddi, E., Cimatti, A., Pozzetti, L., Hoekstra, H. Roettgering, H.J.A., 
Renzini, A., Zamorani, G., Mannucci, F. 2000, A$\&$A, 361, 535
\bibitem []{}Davies, R.L., Sadler, E.M., $\&$ Peletier, R.F., 1993, MNRAS, 262, 650
\bibitem []{}de Freitas Pacheco, J.A., Michard, R., Mohayaee, R. 2003, astro-ph/0301248
\bibitem []{}Djorgovski, S., $\&$ Davis, M. 1987, ApJ, 313, 59
\bibitem []{}Dressler, A., Lynden-Bell, D., Burstein, D., Davies, R.L.,
Faber, S.M., Terlevich, R.J., $\&$ Wegner, G. 1987, ApJ, 313, 42
\bibitem []{}Dressler, A., Oemler, A., Couch, W.J., Smail, I., Ellis, R.S.,
Barger, A., Butcher, H., Poggianti, B.M., $\&$ Sharples, R.M. 1997, ApJ, 490, 577
\bibitem []{}Ellis, R.S., Smail, I., Dressler, A., Couch, W.J., Oemler, A.Jr., Butcher, H., $\&$ Sharples,
R.M., 1997, ApJ, 483, 582
\bibitem []{}Faber, S.M., Burstein, D., $\&$ Dressler, A. 1977, AJ, 82, 941
\bibitem []{}Faber, S.M., Worthey, G., $\&$ Gonzalez, J.J. 1992, in IAU Symp. n.149,
eds. B. Barbuy $\&$ A. Renzini, p. 255
\bibitem []{}Ferreras, I., Charlot, S., $\&$ Silk, J. 1999, ApJ, 521, 81
\bibitem []{}Ferreras, I., $\&$ Silk, J. 2002, MNRAS, 336, 1181
\bibitem []{}Ferreras, I., $\&$ Silk, J. 2003, MNRAS, 344, 455
\bibitem []{}Fran\c cois, P., Matteucci, F., Cayrel, R., Spite, M., Spite, F.,
\& Chiappini, C. 2003, submitted to A\&A
\bibitem []{}Gonzalez, J.J., 1993, PhD thesis, Univ. of California
\bibitem []{}Gonzalez, J.J., $\&$ Gorgas, J. 1996, in ASP Conference Series, 86, Fresh Views of Elliptical Galaxies,
eds. A. Buzzoni, A. Renzini, $\&$ A. Serrano, 225
\bibitem []{}Greggio, L., $\&$ Renzini, A. 1983, A$\&$A, 118, 217
\bibitem []{}Holweger, H. 2001, in Solar and Galactic Composition, ed. R.F. Wimmer-Schweingruber
\bibitem []{}Idiart, T.P, Thevenin, F., $\&$ de Freitas Pacheco, J.A. 1997, AJ, 113, 1066
\bibitem []{}Iwamoto, K., Barchwitx, F., Nomoto, K., Kishimoto, N., Umeda, H., Hix, W.R., Thielemann, F.K. 1999, ApJSS, 125, 439
\bibitem []{}Jaffe, W., 1983, MNRAS, 202, 995
\bibitem []{}Jimenez, R., Padoan, P., Matteucci, F., Heavens, A.F., 1998, MNRAS, 299, 123
\bibitem []{}J\o rgensen, I. 1999, MNRAS, 306, 607 
\bibitem []{}Kauffmann, G., $\&$ Charlot, S. 1998, MNRAS, 294, 705
\bibitem []{}Kauffmann, G., $\&$ White, S.D.M. 1993, MNRAS, 261, 921
\bibitem []{}Kawata, D. 2001, ApJ, 558, 598
\bibitem []{}Kawata, D., $\&$ Gibson, B.K. 2003, MNRAS, 340, 908
\bibitem []{}Kobayashi, C., $\&$ Arimoto, N. 1999, ApJ, 527, 573
\bibitem []{}Kodama, T., $\&$ Arimoto, N. 1997, A$\&$A, 320, 41
\bibitem []{}Kormendy, J., $\&$ Djorgovski, S. 1989, ARA$\&$A, 27, 235
\bibitem []{}Kuntschner, H. 2000, MNRAS, 315, 184
\bibitem []{}Kuntschner, H., Lucey, J.R., Smith, R.J., Hudson, M.J., Davies, R.L.  2001, MNRAS, 323, 615
\bibitem []{}Larson, R.B., 1974, MNRAS, 166, 585
\bibitem []{}Martinelli, A., Matteucci, F., Colafrancesco, S., 1998, MNRAS, 298, 42
\bibitem []{}Matteucci, F., 1992, ApJ, 397, 32
\bibitem []{}Matteucci, F. 1994, A$\&$A, 288, 57
\bibitem []{}Matteucci, F. 2001, The chemical evolution of the Galaxy, Kluwer Academic Publishers, Dordrecht
\bibitem []{}Matteucci, F., $\&$ Fran\c cois, P. 1989, MNRAS, 239, 885
\bibitem []{}Matteucci, F., $\&$ Gibson, B.K.,  1995,  A$\&$A, 304, 11
\bibitem []{}Matteucci, F., $\&$ Greggio, L., 1986, A$\&$A, 154, 279
\bibitem []{}Matteucci, F., $\&$ Pipino, A. 2002, ApJ, 596, 69
\bibitem []{}Matteucci, F., Ponzone, R., Gibson, B.K., 1998, A$\&$A, 335, 855
\bibitem []{}Matteucci, F., $\&$ Tornambe', A., 1987, A$\&$A, 185, 51
\bibitem []{}Mehlert, D., Saglia, R.P., Bender, R., $\&$ Wegner, G. 2000, A$\&$AS, 141, 449
\bibitem []{}Mehlert, D., Thomas, D., Saglia, R.P., Bender, R., $\&$ Wegner, G. 2003, A\&A, 407, 423
\bibitem []{}Menanteau, F., Abraham, R.G., \& Ellis, R.S. 2001a, MNRAS, 322, 1
\bibitem []{}Menanteau, F., Jimenez, R., Matteucci, F. 2001b, ApJ, 562, 23
\bibitem []{}Meynet, G., Maeder, A. 2002, A$\&$A, 381, 25
\bibitem []{}Miyazaki, M., Kodama, T., Okamura, S., et al. 2002, submitted to ApJ, astro-ph/0210509
\bibitem []{}Mobasher, B., Guzman, R., Aragon-Salamanca, A., Zepf, S., 1999, MNRAS, 304, 
225
\bibitem []{}Nipoti, C., Londrillo, P., Ciotti, L. 2002, MNRAS, 332, 901
\bibitem []{}Nipoti, C., Londrillo, P., Ciotti, L. 2003, proceedings of 'The mass of
galaxies at low and high redshift' ESO workshop
\bibitem []{}Nomoto, K., Hashimoto, M., Tsujimoto, T., Thielemann, F.K., Kishimoto, 
N., Kubo, Y., Nakasato, N., 1997, Nuclear Physics A, A621, 467
\bibitem []{}O'Connel, R.W. 1976, ApJ, 206, 370
\bibitem []{}Pagel, B.E.J., $\&$ Patchett, B.E. 1975, MNRAS, 172, 13 
\bibitem []{}Peebles, P.J.E. 2002, in A New Era in Cosmology, ASP
Conference Series, eds. N.Metcalfe $\&$ T.Shanks, in press
\bibitem []{}Peletier, R.F., Davies, R.L., Illingworth, G.D., Davis, L.E., Cawson, M. 1990, AJ, 100, 1091
\bibitem []{}Pipino, A., Matteucci, F., Borgani, S., Biviano, A. 2002, NewA, 7, 227
\bibitem []{}Recchi, S., Matteucci, F., D'Ercole, A., 2001, MNRAS, 322, 800
\bibitem []{}Renzini, A., $\&$ Cimatti, A. 1999, in 'The Hy-Redshift Universe: Galaxy Formation and Evolution at High Redshift',
ASP Conference Proceedings, Vol. 193, A.J. Bunker and W.J.M. van Breugel eds.
\bibitem []{}Renzini, A., $\&$ Ciotti, L.  1983, ApJL, 416, 49
\bibitem []{}Romano, D., Silva, L., Matteucci, F., Danese, L. 2002, MNRAS, 334, 444
\bibitem []{}Rusin, D., Kochanek, C.S., Falco, E.E., Keeton, C.R., McLeod, B.A., Impey, C.D., Lehar, J.,
Munoz, J.A., Peng, C.Y., Rix, H.W. 2003, ApJ, 587, 143
\bibitem []{}Saglia, R.P., Maraston, C., Greggio, L., Bender, R., $\&$ Ziegler, B. 2000, 360, 911
\bibitem []{}Saglia, R.P., Maraston, C., Thomas, D., Bender, R., $\&$ Colless, M. 2002, ApJ, 579, 13
\bibitem []{}Salpeter, E.E., 1955, ApJ, 121, 161
\bibitem []{}Scodeggio, M., Gavazzi, G., Belsole, E., Pierini, D., $\&$ Boselli, A. 1998, MNRAS, 301, 1001
\bibitem []{}Stanford., S.A., Eisenhardt, P.R., $\&$ Dickinson, M. 1998, ApJ, 492, 461
\bibitem []{}Steidel, C.C., Giavalisco, M., Pettini, M., Dickinson, M., $\&$ Adelberger, 
K.L., 1996a, ApJ, 462, L17
\bibitem[]{}Steidel, C.C., Giavalisco, M., Dickinson, M., Adelberger, K.L., 1996b, AJ.,
112, 352
\bibitem []{}Tamura, N., Kobayashi, C., Arimoto, N., Kodama, T., Ohta, K. 2000, AJ, 119, 2134
\bibitem []{}Tantalo, R., Bressan, A., Chiosi, C., 1998a, A$\&$A, 333, 419
\bibitem []{}Tantalo, R., Chiosi, C., Bressan, A., Fagotto, F. 1996, A$\&$A, 311, 361
\bibitem []{}Tantalo, R., Chiosi, C., Bressan, A., Marigo, P., Portinari, L. 1998b, A$\&$A, 335, 823
\bibitem []{}Thielemann, F.K., Nomoto, K., Hashimoto, M. 1996, ApJ, 460, 408 (TNH96)
\bibitem []{}Thomas, D., Greggio, L., $\&$ Bender, R., 1999, MNRAS, 302, 537
\bibitem []{}Thomas, D., $\&$ Kauffmann, G. 1999, in Spectroscopic dating of stars and
galaxies, ASP Conference Series, 192, eds. I. Hubeny, S. Heap, R. Cornett, 261 
\bibitem []{}Thomas, D., Maraston, C., $\&$ Bender, R., 2002, Ap$\&$SS, 281, 371
\bibitem []{}Thomas, D., Maraston, C., $\&$ Bender, R., 2003, MNRAS, 339, 897 
\bibitem []{}Tinsley, B.M., 1980, ApJ, 241, 41
\bibitem []{}Trager, S.C., Faber, S.M., Worthey, G., Gonzalez, J.J., 2000a, AJ, 119, 1654
\bibitem []{}Trager, S.C., Faber, S.M., Worthey, G., Gonzalez, J.J., 2000b, AJ, 120, 165
\bibitem []{}Trager, S.C., Worthey, G., Faber, S.M., Burstein, D., Gonzalez, J.J., 1998 ApJS, 116, 1
\bibitem []{}van de Ven, G., van Dokkum, P.G., Franx, M. 2003, MNRAS, 344, 924
\bibitem []{}van den Hoek, L.B., Groenewegen, M.A.T. 1997, A$\&$AS, 123, 305
\bibitem []{}van Dokkum, P.G., $\&$ Franx, M. 1996, MNRAS, 281, 985
\bibitem []{}van Dokkum, P.G., Franx, M., Kelson., D.D., Illingworth, G.D., Fisher, D., Fabricant, D. 1998, ApJ, 500, 714
\bibitem []{}Weiss, A., Peletier, R.F., Matteucci, F. 1995, A$\&$A, 296, 73
\bibitem []{}Whelan, J., Iben, I. Jr. 1973, ApJ, 186, 1007
\bibitem []{}White, S.D.M., $\&$ Rees, M.J., 1978, MNRAS, 183, 341
\bibitem []{}Woosley, S.E., $\&$ Weaver, T.A., 1995, ApJS, 101, 181 (WW95)
\bibitem []{}Worthey, G. 1994, ApJS, 95, 107
\bibitem []{}Worthey, G. 1998, PASP, 110, 888
\bibitem []{}Worthey, G., $\&$ Collobert, M. 2003, ApJ, 586, 17
\bibitem []{}Worthey, G., Dorman, B., Jones, L.A. 1996, AJ, 112, 948 
\bibitem []{}Worthey, G., Faber, S.M., $\&$ Gonzalez, J.J. 1992, ApJ, 398, 69


\end{thebibliography}
\end{document}